\documentclass{ieeetmlcn}

\usepackage{cite}
\usepackage{multirow}
\usepackage{amsmath,amssymb,amsfonts}
\usepackage[ruled,vlined,linesnumbered]{algorithm2e}
\usepackage{graphicx,color}
\usepackage{textcomp}
\usepackage{xcolor}
\usepackage{subcaption}
\usepackage{bm}
\usepackage{tabularray}
\usepackage{longtable}
\usepackage{anyfontsize}
\usepackage[font=small]{caption}
\usepackage{hyperref}
\hypersetup{hidelinks=true}

\def\BibTeX{{\rm B\kern-.05em{\sc i\kern-.025em b}\kern-.08em
    T\kern-.1667em\lower.7ex\hbox{E}\kern-.125emX}}

\AtBeginDocument{%
  \definecolor{tmlcncolor}{cmyk}{0.93,0.59,0.15,0.02}%
  \definecolor{NavyBlue}{RGB}{0,86,125}}

\def\OJlogo{\vspace{-4pt}\includegraphics[height=18pt]{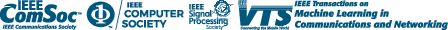}}
\def\seclogo{\vspace{10pt}\includegraphics[height=18pt]{new_logo}}
\def\authorrefmark#1{\ensuremath{^{\textbf{#1}}}}

\SetCommentSty{italic}
\SetKwComment{Commentxs}{$\triangleright$\ }{}
\let\oldnl\nl
\newcommand{\nonl}{\renewcommand{\nl}{\let\nl\oldnl}}

\begin{document}

\receiveddate{XX Month, XXXX}
\reviseddate{XX Month, XXXX}
\accepteddate{XX Month, XXXX}
\publisheddate{XX Month, XXXX}
\currentdate{XX Month, XXXX}
\doiinfo{TMLCN.XXXX.XXXXXXX}

\markboth{}{Ghassemi {et al.}}

\title{Dual-Transformer Aided Hierarchical Deep Reinforcement Learning
       for Robust RIS-Assisted Near-Field Communications}

\author{Mohammad Ghassemi\authorrefmark{1},
  Han Zhang\authorrefmark{1},
  Ali Afana\authorrefmark{2},
  Akram Bin Sediq\authorrefmark{2}, and
  Melike Erol-Kantarci\authorrefmark{1},
  \textit{Fellow}, IEEE}

\affil{School of Electrical Engineering and Computer Science, University of Ottawa, Ottawa, ON K1N 6N5, Canada}
\affil{Ericsson, Ottawa, ON K2K 2V6, Canada}

\corresp{Corresponding author: M. EROL-KANTARCI (melike.erolkantarci@uottawa.ca).}

\authornote{This work was supported by the Mathematics of Information Technology and Complex Systems (MITACS) and Ericsson Canada and in part by the NSERC Canada Research Chairs Program.}

\begin{abstract}
The deployment of extremely large aperture arrays (ELAAs) in sixth-generation (6G) networks is expected to shift communications into the near-field regime, where spherical-wave propagation enables distance-aware beamfocusing but remains highly vulnerable to physical blockages that cause non-line-of-sight (NLoS) conditions.
To resolve this inherent vulnerability, reconfigurable intelligent surfaces (RIS) can be utilized to circumvent these blockages and effectively establish reliable NLoS communication links.
In envisioned deployment scenarios, accurately acquiring instantaneous CSI and predicting sudden blockages is profoundly challenging due to the prohibitive pilot overhead associated with massive passive arrays and the unpredictable mobility of environmental scatterers.
To address this, we propose the Dual-Transformer Hierarchical Deep Reinforcement Learning (DT-HDRL) framework, which integrates two specialized transformer models with a two-timescale hierarchical control agent.
The first transformer integrates a ray-tracing digital twin prior with distance-aware geometric correction features to yield rapid and precise CSI estimates, while a complementary vision transformer (ViT) processes sequential camera frames to forecast impending blockages prior to link degradation.
These predictive outputs are then fed directly into the hierarchical control agent.
Within this architecture, a high-level controller processes the slow-timescale blockage predictions to jointly dictate the user transmission path (line-of-sight (LoS) or RIS-assisted NLoS) and RIS active/sleep scheduling, whereas a low-level controller employs the fast-timescale CSI estimates to perform joint base station (BS) beamfocusing and RIS phase-shift optimization.
Simulation results demonstrate an 18.0\% sum spectral efficiency (SE) improvement and a ViT blockage prediction F1-score of 0.92 with a 769~ms advance warning window.
\end{abstract}


\maketitle

\section{INTRODUCTION}
\label{sec:intro}

\IEEEPARstart{S}{ixth-generation} (6G) wireless networks are expected to deliver unprecedented data rates, ultra-low latency, and massive connectivity~\cite{saad2020vision}. A key enabler envisioned for 6G is the deployment of extremely large aperture arrays (ELAAs) expected to provide the spatial multiplexing gain required to meet these targets~\cite{rappaport2019wireless}. By massively increasing the number of antennas, propagation dynamics in such ELAA-based systems are expected to shift into the near-field communication (NFC) regime, requiring spherical-wave channel models that jointly account for angle and distance~\cite{cui2023near,wei2022codebook}. To exploit these spherical wavefronts, the network would need to transition from traditional beamforming to beamfocusing, enabling the base station to concentrate signal energy at specific physical locations~\cite{bjornson2020power}.

Despite the theoretical advantages of beamfocusing, the realization of NFC faces two primary physical challenges. First, high-frequency bands suffer from severe path loss and are easily blocked by physical objects~\cite{rangan2014millimeter}. Second, spherical wavefronts introduce significant nonlinearity into CSI estimation~\cite{cui2023near,lei2023channel}. Consequently, conventional pilot-based CSI acquisition incurs massive computational costs when attempting to extract accurate high-dimensional CSI~\cite{heath2016overview}, creating a severe bottleneck for beamfocusing~\cite{ma2020sparse}. To bypass this pilot overhead, digital twins offer a compelling alternative by utilizing known environmental geometry and ray tracing to simulate the channel directly~\cite{jiang2025learnable}. However, pure digital twin models are often inaccurate in dynamic environments because they are unable to perfectly predict random, dynamic scattering. To overcome this limitation, a promising approach is to learn a correction over a ray-tracing digital twin prior, fusing geometric features to recover the stochastic fading components that pure digital twin models fail to capture.

Furthermore, to actively mitigate severe path loss and physical obstructions, reconfigurable intelligent surfaces (RIS) have emerged as a highly effective solution~\cite{direnzo2020smart}. An RIS comprises a massive array of passive elements that intelligently adjust the phase of incident signals to establish virtual NLoS links whenever direct LoS paths become blocked~\cite{wu2020towards,zhang2022beam}. However, integrating these surfaces introduces two additional operational difficulties. First, it intensifies the channel estimation burden by creating a complex cascaded link from the base station (BS) through the RIS to the user equipment (UE) that must be continuously tracked. Second, maintaining a reliable connection requires the network to proactively reconfigure the RIS before a blockage actually occurs. Because rapid channel tracking and slower blockage prediction operate on fundamentally different timescales, managing this architecture necessitates a hierarchical control approach.

\subsection{RELATED WORK}

\subsubsection{NFC Channel Modeling and Estimation}
Accurately acquiring CSI for near-field beamfocusing and RIS systems has become a primary focus in recent literature~\cite{cui2022channel}. Traditional compressive sensing algorithms suffer severe performance degradation in NFC due to spherical energy dispersion. While deep learning techniques utilizing convolutional neural networks (CNNs) can reconstruct cascaded RIS channels~\cite{elbir2020deep}, localized receptive fields limit their ability to capture spatial non-stationarity. More recently, spatial-attention mechanisms and transformer networks have been introduced to improve near-field multiple-input multiple-output (MIMO) estimation~\cite{wang2022transformer,zhu2025spatial}. Despite these advancements, purely data-driven models struggle to generalize across sudden physical blockages in dynamic environments. Furthermore, they fail to accurately reconstruct stochastic NLoS fading components without prohibitive training overhead~\cite{wang2024wideband}, leaving the efficient resolution of unpredictable stochastic variations an open challenge.

\subsubsection{Blockage Prediction}
Beyond channel estimation, link blockage poses a critical challenge for NFC systems. Traditional approaches such as relay switching are reactive, responding only after a blockage has already occurred. Recent research has shifted toward proactive strategies using multi-modal sensors~\cite{alkhateeb2018machine,charan2021vision}, and methods fusing multi-modal data have shown promise in building environment-aware networks~\cite{ghassemi2024multi}. Vision transformers (ViTs) demonstrate strong performance in blockage prediction by processing sequential image data to identify motion patterns that indicate future blockages, offering superior generalization compared to classical signal processing methods~\cite{yang2023hierarchical}. Specifically, computer vision models have been shown to successfully leverage camera feeds to anticipate moving blockages prior to signal degradation~\cite{charan2022computer,ghassemi2025generative}.

\subsubsection{Hierarchical Control in Wireless Networks}
Reinforcement learning (RL) has been widely applied to RIS optimization~\cite{ghassemi2025foundation}. However, standard RL agents suffer from the curse of dimensionality when controlling ELAAs across fundamentally mismatched timescales. To mitigate this computational complexity, hierarchical architectures decompose the problem into manageable sub-tasks~\cite{nguyen2022ris,zhi2022two}. This decomposition is critical for RIS systems, allowing rapid beamfocusing adaptation to fast fading alongside slower path selection decisions. For example, recent frameworks separate high-level macro-control from low-level RIS adjustments to improve convergence~\cite{zhou2022hierarchical}. Despite these structural benefits, such models rely strictly on reactive channel metrics rather than predictive environmental data. Consequently, these conventional agents are unable to proactively anticipate sudden blockages and struggle to process the complex geometric features required for near-field environments. In contrast, DT-HDRL differs from these reactive hierarchical approaches by incorporating camera-based visual blockage prediction, which enables proactive path switching and RIS scheduling before link degradation rather than responding to channel metrics after a blockage has already occurred.

\subsection{MOTIVATIONS AND CONTRIBUTIONS}

To resolve these challenges, we propose the Dual-Transformer Hierarchical Deep Reinforcement Learning (DT-HDRL) framework. The primary contributions of this paper are summarized as follows.

\begin{itemize}

  \item \textbf{Dual-Modal Transformer Design.}
    We design two specialized transformer models tailored for the near-field regime.
    The first fuses a ray-tracing digital twin prior with distance-aware geometric correction features to estimate the effective near-field channel, thereby recovering stochastic fading components omitted by pure digital twin models and achieving a 73\% lower mean squared error (MSE) than CNN baselines.
    The second model utilizes a ViT to analyze sequential visual data, predicting future blockage events with a 769~ms advance warning window and an F1-score of 0.92.

  \item \textbf{Dual-Timescale Hierarchical Control.}
    We introduce the DT-HDRL framework to align with environmental timescales.
    A high-level controller processes slow-timescale blockage predictions to jointly dictate the user transmission path (LoS or RIS-assisted NLoS) and RIS active/sleep scheduling, whereas a low-level controller employs fast-timescale CSI estimates to perform joint base station (BS) beamfocusing and RIS phase-shift optimization.
    This hierarchical structure maximizes the sum spectral efficiency (SE) based on the established routing and scheduling strategy, achieving an 18.0\% SE improvement over single-timescale baselines.

  \item \textbf{Robustness and Performance Analysis.}
    We validate the DT-HDRL framework using the DeepVerse 6G Carla-Town5 suburban outdoor dataset.
    We systematically evaluate the framework to isolate the individual performance gains provided by the distance-aware CSI transformer, the ViT blockage predictor, and the hierarchical architecture.
    Additionally, we analyze the effects of imperfect CSI and UE location errors to demonstrate system robustness under channel impairment conditions.

\end{itemize}

The remainder of this paper is organized as follows. Section~\ref{sec:model} introduces the system model and problem formulation. Section~\ref{sec:framework} details the proposed DT-HDRL framework, comprising the dual-transformer estimators and the hierarchical control agent. Section~\ref{sec:results} presents numerical results and performance analysis. Section~\ref{sec:conclusion} concludes the paper.

\section{SYSTEM MODEL AND PROBLEM FORMULATION}
\label{sec:model}

In this section, we develop the mathematical model for RIS-assisted NFC and formulate the joint beamfocusing and phase-shift optimization problem. To ensure consistency throughout the paper, we adopt the following notation. Bold lowercase letters denote vectors, bold uppercase letters denote matrices, and italic letters denote scalars. The macro-step and micro-step indices are denoted by $t'$ and $\tau$, respectively, and the effective channel is denoted by $\mathbf{h}_{\text{eff},k}$.

We first describe the BS and RIS configurations, then derive the near-field channel model, and finally state the optimization problem that DT-HDRL is designed to solve.

\subsection{SYSTEM MODEL}

We consider a downlink NFC system comprising a BS equipped with an ELAA, a set $\mathcal{R}$ of passive RIS units, and $K$ single-antenna UEs indexed by $\mathcal{K} = \{1, \ldots, K\}$. Each UE $k$ is located at $\mathbf{u}_k = [x_k, y_k, z_k]^T$ in Cartesian coordinates.

\subsubsection{Base Station Configuration}
The BS is equipped with an ELAA configured as a uniform planar array (UPA) of $N = N_y \times N_z$ antennas aligned along the $y$- and $z$-axes, with its center at the origin $\mathbf{u}_{\text{BS}} = [0, 0, 0]^T$. Using linear indexing $n \in \{1, \ldots, N\}$, the position of the $n$-th antenna element relative to the array center is:
\begin{equation}
  \mathbf{p}_{n}^{\text{BS}} = [0,\; n_y d_{\text{BS}},\; n_z d_{\text{BS}}]^T,
\end{equation}
where $d_{\text{BS}}$ is the inter-element spacing, and $n_y \in \mathcal{N}_y$, $n_z \in \mathcal{N}_z$ are local indices along the $y$- and $z$-axes, respectively. The notation $[\cdot]^T$ denotes the transpose. To center the array at $\mathbf{u}_{\text{BS}}$, the index sets are $\mathcal{N}_y = \{-\frac{N_y-1}{2}, \ldots, \frac{N_y-1}{2}\}$ and $\mathcal{N}_z = \{-\frac{N_z-1}{2}, \ldots, \frac{N_z-1}{2}\}$.\footnote{We assume a fully digital beamfocusing architecture at the BS for conceptual clarity. Extension to hybrid beamfocusing architectures and explicit interference management strategies are left for future work.}

\subsubsection{RIS Configuration}
Each RIS $r \in \mathcal{R}$ is a UPA composed of $M = M_y \times M_z$ passive metasurface elements. The center of RIS $r$ is located at $\mathbf{u}_{\text{RIS},r}$, with RIS~1 deployed on a building facade at $\mathbf{u}_{\text{RIS},1}$ and RIS~2 on an opposing facade at $\mathbf{u}_{\text{RIS},2}$, providing complementary NLoS coverage of the road segment. This placement ensures that blocked UEs in the road corridor always have at least one RIS with a favorable reflection angle. The position of the $m$-th element of RIS $r$ relative to its center is:
\begin{equation}
  \mathbf{p}_{m}^{\text{RIS},r} = [0,\; m_y d_{\text{RIS}},\; m_z d_{\text{RIS}}]^T,
\end{equation}
where $d_{\text{RIS}}$ is the element spacing, and the indices $m_y$, $m_z$ are drawn from centered index sets $\mathcal{M}_y$ and $\mathcal{M}_z$ defined analogously to those of the BS. The absolute global coordinate of the $m$-th element of RIS $r$ is $\mathbf{u}_{\text{RIS},r} + \mathbf{p}_{m}^{\text{RIS},r}$. Each RIS $r$ is characterized by its reflection coefficient matrix $\bm{\Theta}_r = \text{diag}(e^{j\phi_{1,r}}, \ldots, e^{j\phi_{M,r}}) \in \mathbb{C}^{M \times M}$, where $\phi_{m,r} \in [0, 2\pi)$ is the controllable phase shift of the $m$-th element of RIS $r$. Each RIS $r$ additionally operates under an active/sleep scheduling policy. A binary indicator $a_r \in \{0, 1\}$ determines whether RIS $r$ is active ($a_r = 1$, reflecting signals) or sleeping ($a_r = 0$, powered off). This scheduling degree of freedom allows the controller to selectively activate only those RIS units that provide a meaningful SNR gain, thereby avoiding unnecessary hardware power consumption.

\subsubsection{NFC Spherical Wave Channel Model}
We adopt the spherical wave model, which accounts for both phase variations and distance-dependent path loss across the full array aperture~\cite{bjornson2020power}. The system operates in both LoS (direct BS-to-UE) and NLoS (BS-to-RIS-to-UE) propagation modes in the RIS-assisted NFC. The ray-tracing engine used to generate the DeepVerse 6G dataset captures reflections and scattering up to 5th order, so the digital twin prior fed to the CSI transformer inherently encodes rich NLoS multipath components on all links including the cascaded BS-to-RIS-to-UE paths. The geometric channel equations below therefore describe the dominant propagation structure. In practice, the received signal is a superposition of multipath components with small-scale fading. To account for this, the ground-truth effective channel $\mathbf{h}_{\text{eff},k}$ includes a stochastic residual that the CSI transformer is trained to recover from the deterministic digital-twin prior.

The geometric correction branch of the transformer recovers the stochastic residual, including small-scale fading and diffuse scattering.

\textbf{1) BS-to-UE Direct Channel.} The distance from the $n$-th BS antenna to UE $k$ is $d_{k,n}^{\text{BD}} = \|\mathbf{u}_k - \mathbf{p}_n^{\text{BS}}\|$, where $\|\cdot\|$ denotes the Euclidean norm. The channel coefficient from the $n$-th BS antenna to UE $k$ is modeled as~\cite{bjornson2020power}:
\begin{equation}
  [\mathbf{h}_{\text{BD},k}]_{n} =
    \sqrt{\eta_{k,n}^{\text{BD}}} \cdot
    e^{-j\frac{2\pi}{\lambda}d_{k,n}^{\text{BD}}},
  \label{eq:channel_bd_element}
\end{equation}
where $\lambda$ is the carrier wavelength and $\eta_{k,n}^{\text{BD}} = \bigl(\frac{\lambda}{4\pi d_{k,n}^{\text{BD}}}\bigr)^2$ is the free-space path loss. The full channel vector is $\mathbf{h}_{\text{BD},k} \in \mathbb{C}^{N \times 1}$.

\textbf{2) RIS-to-UE Channel.} For RIS $r \in \mathcal{R}$, the distance from the $m$-th element of RIS $r$ to UE $k$ is $d_{k,m,r}^{\text{RD}} = \|\mathbf{u}_k - (\mathbf{u}_{\text{RIS},r} + \mathbf{p}_m^{\text{RIS},r})\|$. The $m$-th element of the channel vector $\mathbf{h}_{\text{RD},k,r} \in \mathbb{C}^{M \times 1}$ is:
\begin{equation}
  [\mathbf{h}_{\text{RD},k,r}]_{m} =
    \sqrt{\eta_{k,m,r}^{\text{RD}}} \cdot
    e^{-j\frac{2\pi}{\lambda}d_{k,m,r}^{\text{RD}}},
  \label{eq:channel_rd_element}
\end{equation}
where $\eta_{k,m,r}^{\text{RD}} = \bigl(\frac{\lambda}{4\pi d_{k,m,r}^{\text{RD}}}\bigr)^2$ is the path loss for the link from RIS~$r$ to UE~$k$.

\textbf{3) BS-to-RIS Channel.} The channel between the BS and RIS $r$ is $\mathbf{G}_{\text{BR},r} \in \mathbb{C}^{M \times N}$, where the $(m,n)$-th entry is:
\begin{equation}
  [\mathbf{G}_{\text{BR},r}]_{m,n} =
    \sqrt{\eta_{m,n,r}^{\text{BR}}} \cdot
    e^{-j\frac{2\pi}{\lambda}d_{m,n,r}^{\text{BR}}},
  \label{eq:channel_br_element}
\end{equation}
where $d_{m,n,r}^{\text{BR}} = \|(\mathbf{u}_{\text{RIS},r} + \mathbf{p}_m^{\text{RIS},r}) - \mathbf{p}_n^{\text{BS}}\|$ and $\eta_{m,n,r}^{\text{BR}} = \bigl(\frac{\lambda}{4\pi d_{m,n,r}^{\text{BR}}}\bigr)^2$.

\subsection{NFC PROPAGATION ANALYSIS}

The boundary between near-field and far-field propagation is defined by the Rayleigh distance~\cite{bjornson2020power,cui2023near}:
\begin{equation}
  Z_R = \frac{2 D^2}{\lambda},
\end{equation}
where $D$ is the maximum array aperture dimension. For the baseline configuration of $N = 1024$ antennas arranged as a $32 \times 32$ UPA at $d_{\text{BS}} = \lambda/2$ and $f_c = 3.5$~GHz ($\lambda \approx 85.7$~mm), the linear aperture along each axis is $D_y = D_z \approx 1.33$~m, and the effective diagonal aperture is $D \approx 1.88$~m, yielding $Z_R \approx 82$~m. Both RIS units (with $d_{\text{RIS}} = \lambda/2$) and vehicular UEs on the closest road segments therefore operate within the near-field region, validating the spherical-wave model throughout the deployment.

To quantify the error introduced by the planar-wave approximation, consider a user at radial distance $\varrho$ and angle $\theta$ from the array boresight, and let $q_n$ denote the scalar position of the $n$-th element along the aperture. Note that $\varrho$ is used here for the user-to-array radial distance to avoid notational conflict with the RIS unit index $r \in \mathcal{R}$ used throughout the paper. Under the planar-wave assumption, the phase is $\phi_{\text{planar}}(n) = \frac{2\pi}{\lambda} q_n \sin(\theta)$, while the exact spherical phase is~\cite{bjornson2020power}:
\begin{equation}
  \phi_{\text{spherical}}(n) =
    \frac{2\pi}{\lambda}
    \left( \sqrt{\varrho^2 + q_n^2 - 2\varrho q_n\sin(\theta)}
    - \varrho \right).
\end{equation}
The phase error is defined as the difference between the exact spherical phase and the planar-wave approximation, $\Delta\phi(n) = \phi_{\text{spherical}}(n) - \phi_{\text{planar}}(n)$. A second-order Taylor expansion gives:
\begin{equation}
  \Delta \phi(n) \approx \frac{\pi q_n^2 \cos^2(\theta)}{\lambda \varrho}.
\end{equation}
The remainder of this expansion is bounded by $\frac{\pi |q_n^3 \sin\theta|}{\lambda \varrho^2} + \mathcal{O}(q_n^4/\varrho^3)$, which falls below $0.11$~rad for the largest array dimension ($q_n \approx 0.66$~m) and the minimum user distance ($\varrho \ge 10$~m) in our setup, confirming sufficient accuracy for SINR computation. This bound corresponds to a maximum SNR deviation of less than $0.05$~dB.

Since our study focuses on narrowband transmission at $f_c = 3.5$~GHz, beam squint does not arise. Its mitigation in wideband orthogonal frequency division multiplexing (OFDM) scenarios is left for future work. In the proposed framework, the CSI transformer uses the exact Cartesian distances as geometric correction features on top of the ray-tracing digital twin prior, enabling precise energy focusing without requiring explicit phase compensation. While DT-HDRL is designed for near-field beamfocusing, the hierarchical control structure and the transformer-based CSI estimator remain applicable to far-field scenarios by replacing the spherical-wave geometric features with planar-wave angles in the input sequence (Eq.~(\ref{eq:csi_input})).

\subsection{PROBLEM FORMULATION}

A binary blockage indicator $b_k \in \{0, 1\}$ is defined for each UE $k$, where $b_k = 1$ indicates an available LoS link and $b_k = 0$ indicates a blocked link that must be served via the RIS-assisted NLoS path. Let $\mathbf{W} = [\mathbf{w}_1, \ldots, \mathbf{w}_K] \in \mathbb{C}^{N \times K}$ be the BS beamfocusing matrix and $\mathbf{s} = [s_1, \ldots, s_K]^T \in \mathbb{C}^{K \times 1}$ be the symbol vector with $\mathbb{E}[|s_k|^2] = 1$ for all $k \in \mathcal{K}$. Multi-user interference is handled by the beamfocusing design. No successive interference cancellation (SIC) is assumed, so the SINR expression follows conventional linear precoding. The received signal at UE $k$ is:
\begin{equation}
  y_k = \mathbf{h}_{\text{eff},k}^H \mathbf{w}_k s_k
        + \sum_{i \neq k} \mathbf{h}_{\text{eff},k}^H \mathbf{w}_i s_i
        + n_k,
  \label{eq:received_unified}
\end{equation}
where $n_k \sim \mathcal{CN}(0, \sigma^2)$ is additive white Gaussian noise (AWGN) at UE $k$ with power $\sigma^2$, and the effective channel vector is:
\begin{equation}
  \mathbf{h}_{\text{eff},k}^H =
    b_k \mathbf{h}_{\text{BD},k}^H
    + (1-b_k) \sum_{r \in \mathcal{R}}
      a_r \mathbf{h}_{\text{RD},k,r}^H \bm{\Theta}_{r}
      \mathbf{G}_{\text{BR},r},
  \label{eq:effective_channel}
\end{equation}
where $a_r \in \{0,1\}$ is the active/sleep indicator for RIS $r$. An active RIS ($a_r=1$) contributes its reflected signal to the NLoS path, while a sleeping RIS ($a_r=0$) is powered off and contributes nothing. When the direct LoS link for a given UE is blocked ($b_k = 0$), the placement described in Section~\ref{sec:model}-A ensures that at least one RIS provides a connection. Consequently, the active/sleep indicators $\{a_r\}$ and phase-shift matrices $\{\bm{\Theta}_r\}$ for the active units are jointly optimized to maximize the combined reflected power. The signal-to-interference-plus-noise ratio (SINR) at UE $k$ is:
\begin{equation}
  \text{SINR}_k = \frac{%
    |\mathbf{h}_{\text{eff},k}^H \mathbf{w}_k|^2}{%
    \sum_{i \neq k} |\mathbf{h}_{\text{eff},k}^H \mathbf{w}_i|^2
    + \sigma^2},
  \label{eq:sinr}
\end{equation}
and the SE for UE $k$ is:
\begin{equation}
  \text{SE}_{k} = \log_2(1 + \text{SINR}_{k})
  \quad \text{(bps/Hz)}.
  \label{eq:se_definition}
\end{equation}

In practice, the controller relies on the predicted blockage state $\hat{b}_k$ from the ViT and the estimated channel $\hat{\mathbf{h}}_{\text{eff},k}$ from the CSI transformer rather than their perfect values. Consequently, the SINR and SE computed from these estimated quantities are denoted $\widehat{\text{SINR}}_k$ and $\widehat{\text{SE}}_k$, respectively. The joint optimization problem is:
\begin{equation}
  \begin{aligned}
    \label{eq:overall_opt}
\max_{\mathbf{W},\, \{a_r\},\, \{\bm{\Theta}_r\}} \quad & \sum_{k=1}^{K} \widehat{\text{SE}}_k \\ \text{s.t.} \quad & \|\mathbf{W}\|_F^2 \le P_{\text{max}}, \quad \text{(C1)} \\ & \phi_{m,r} \in [0, 2\pi),                   \\ & \quad \forall m \in \{1, \ldots, M\},\; r \in \mathcal{R},  \quad \text{(C2)} \\ & a_r \in \{0, 1\}, \quad \forall r \in \mathcal{R}, \quad \text{(C3)} \\ & \textstyle\sum_{r \in \mathcal{R}} a_r \ge \max_{k \in \mathcal{K}}(1-b_k), \quad \text{(C4)} \\ & \widehat{\text{SE}}_k \ge \text{SE}_{\text{min}}, \quad \forall k \in \mathcal{K}. \quad \text{(C5)}
  \end{aligned}
\end{equation}
Constraint~(C1) enforces the BS transmit power limit. Constraint~(C2) restricts each RIS phase shift to the admissible continuous interval, where each $\phi_{m,r}$ parameterizes the corresponding diagonal entry of $\bm{\Theta}_r$. Although the phase constraint $\phi_{m,r} \in [0, 2\pi)$ is technically redundant for the diagonal phase-shift matrix $\bm{\Theta}_r$ because $e^{j\phi}$ is $2\pi$-periodic, we enforce this bound during DRL training to stabilize the continuous action space and prevent unbounded phase drift in the DDPG policy.

Constraint~(C3) defines the binary active/sleep scheduling variable for each RIS unit. Constraint~(C4) ensures that at least one RIS remains active whenever any UE is in NLoS conditions, while permitting all RIS units to sleep during full LoS operation. Constraint~(C5) guarantees the minimum per-user quality of service (QoS).

Problem~\eqref{eq:overall_opt} is highly non-convex due to the multiplicative coupling of $\mathbf{W}$, $\{a_r\}$, and $\{\bm{\Theta}_r\}$ within the effective channel. Consequently, solving this formulation via exhaustive search introduces exponential complexity proportional to $2^{|\mathcal{R}|} \times 2M$. Furthermore, alternating optimization (AO) methods converge too slowly for real-time applications. This computational bottleneck is compounded by an inherent timescale disparity, as physical blockages evolve much slower than fast fading channel conditions, making it highly inefficient to update both CSI and blockage predictions at identical rates. Together, these operational challenges necessitate the hierarchical decomposition proposed in the next section.

\section{PROPOSED DUAL-TRANSFORMER HIERARCHICAL FRAMEWORK}
\label{sec:framework}

\begin{figure*}[!t]
  \centering
  \includegraphics[width=0.9\linewidth]{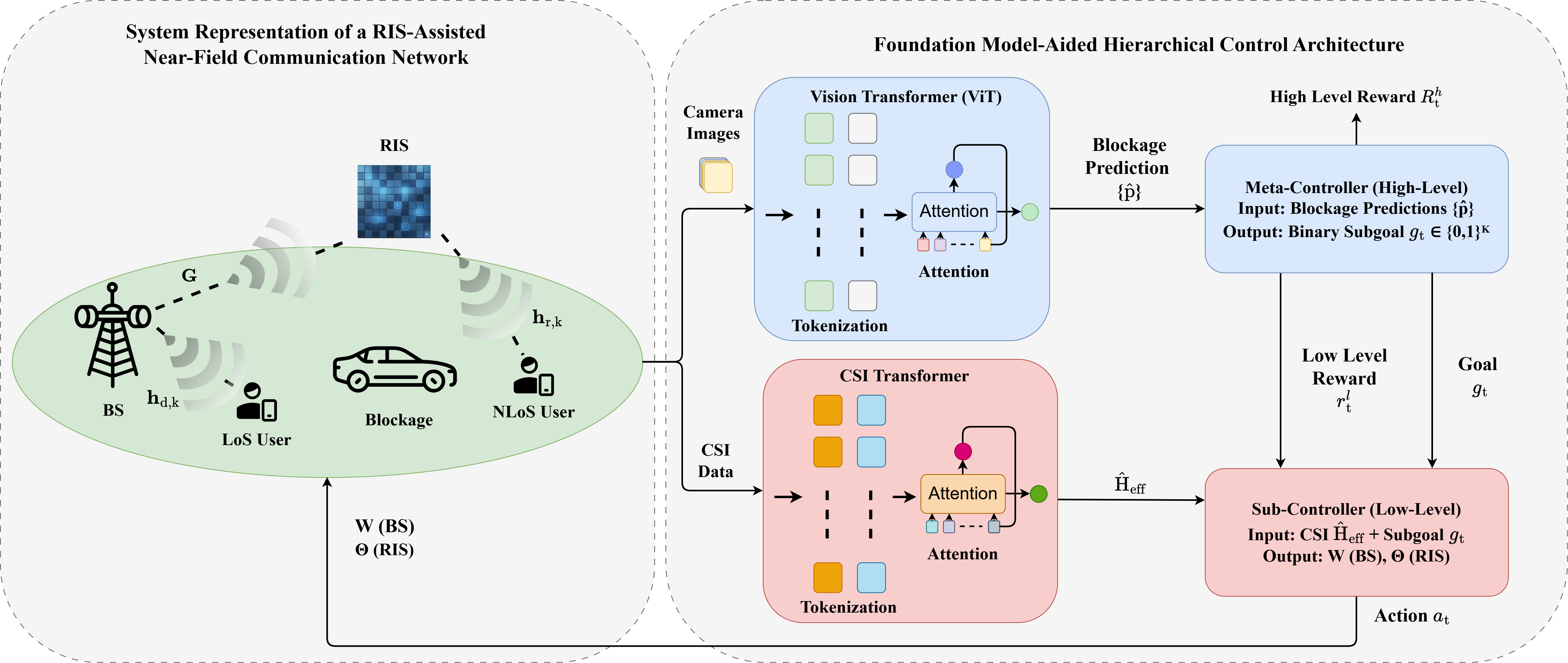}
  \caption{DT-HDRL framework overview.
    \textit{Left}: Physical system representation of the RIS-assisted near-field communication network, depicting the BS equipped with an ELAA, two RIS units, a blockage vehicle, a LoS user, and an NLoS user together with the key channel components $\mathbf{h}_{\text{BD},k}$ and $\mathbf{G}_{\text{BR},r}$.
    \textit{Right}: DT-HDRL control architecture comprising the ViT blockage predictor driven by camera images, the CSI transformer driven by ray-tracing channel data, and the two-timescale hierarchical agent, in which the meta-controller (slow timescale $t'$) performs joint path selection and RIS active/sleep scheduling while the sub-controller (fast timescale $\tau$) performs BS beamfocusing and RIS phase-shift optimization.}
  \label{fig:framework_overview}
\end{figure*}

This section presents DT-HDRL, which addresses Problem~\eqref{eq:overall_opt} through three components: a transformer-based CSI estimator (Section~\ref{sec:framework}-A), a ViT-based blockage predictor (Section~\ref{sec:framework}-B), and an HDRL agent (Section~\ref{sec:framework}-C). Fig.~\ref{fig:framework_overview} shows the overall architecture.

\subsection{TRANSFORMER-BASED NFC CSI ESTIMATION}

The CSI transformer estimates the effective near-field channel $\mathbf{h}_{\text{eff},k}$ for each user $k$. Rather than relying on geometric features alone, the model fuses a ray-tracing digital twin prior with per-antenna geometric correction features. The digital twin prior encodes the dominant deterministic propagation geometry, whereas the correction features enable the transformer to recover stochastic fading and higher-order NLoS components omitted by the purely deterministic prior.

\subsubsection{Input Feature Extraction}
The DeepVerse 6G Carla-Town5 dataset provides ray-tracing channel vectors generated with up to 5th-order reflections. The digital twin prior $\mathbf{h}_k^{\text{DT}}$ is queried from a pre-computed ray-tracing database indexed by the UE location. This database is generated offline for the deployment environment and cached at the edge server, so the online query latency is negligible compared to the macro-step interval.

These channel vectors are assembled into the digital twin prior $\mathbf{h}_k^{\text{DT}} \in \mathbb{C}^{N \times 1}$, which captures the dominant geometric propagation structure for UE $k$ conditioned on its position. Since each UE is equipped with a single antenna and the BS has $N$ antennas, the prior is an $N$-dimensional complex vector representing the BS-side channel response; if the raw ray-tracing output has a different dimension, a learnable linear projection maps it to $\mathbb{C}^{N \times 1}$ before fusion. However, ray-tracing alone is unable to represent small-scale stochastic fading, diffuse scattering, or dynamic blockage events. To recover these missing components, the CSI transformer fuses the digital twin prior with per-antenna distance-aware, geometric-correction features computed analytically from the known UE location. The complete input sequence for UE $k$ is formed by concatenating, for each BS antenna element $n$, the real and imaginary parts of the digital twin prior with the three geometric correction features:
\begin{equation}
  \mathbf{x}_k = \left\{\left[
    [\mathbf{h}_k^{\text{DT}}]_n^{\text{re}},\;
    [\mathbf{h}_k^{\text{DT}}]_n^{\text{im}},\;
    d_{k,n}^{\text{BD}},\;
    \vartheta_{k,n}^{\text{BD}},\;
    \psi_{k,n}^{\text{BD}}
  \right]\right\}_{n=1}^N \in \mathbb{R}^{N \times 5},
  \label{eq:csi_input}
\end{equation}
where $[\mathbf{h}_k^{\text{DT}}]_n^{\text{re}}$ and $[\mathbf{h}_k^{\text{DT}}]_n^{\text{im}}$ denote the real and imaginary parts of the $n$-th element of the digital twin prior. The three geometric correction features for antenna element $n$ are defined as follows. The distance $d_{k,n}^{\text{BD}}$ matches the BS-to-UE channel notation from Section~\ref{sec:model}. The elevation angle $\vartheta_{k,n}^{\text{BD}} = \arcsin\!\bigl(\frac{z_k - z_n^{\text{BS}}}{d_{k,n}^{\text{BD}}}\bigr)$ is in radians. The azimuth angle $\psi_{k,n}^{\text{BD}} = \arctan\!\bigl(\frac{y_k - y_n^{\text{BS}}}{x_k - x_n^{\text{BS}}}\bigr)$ is also in radians. These five features form a complementary representation in which the digital twin prior provides the dominant channel structure while the distance feature $d_{k,n}^{\text{BD}}$ encodes spherical wavefront curvature, enabling the model to distinguish users at the same angle but at different depths, a critical capability in the near-field regime. We assume that the BS has an estimate of UE locations $\mathbf{u}_k$, obtained, for example, from 5G NR positioning reference signals. These locations are obtained via 5G NR positioning reference signals (PRS), which provide sub-meter accuracy in outdoor environments. As validated in Section~\ref{sec:ablation}, even with a location error standard deviation of $\sigma_{\text{loc}} = 0.5$~m, the sum spectral efficiency degrades by only 4.2\%, confirming that sub-meter positioning is sufficient for the proposed geometric correction.

\subsubsection{Feature Embedding and Positional Encoding}
The five-dimensional input $\mathbf{x}_k \in \mathbb{R}^{N \times 5}$ is projected into a latent space of dimension $d_{\text{model}}$ via a learnable linear layer: $\mathbf{H}^{(0)} = \text{Linear}(\mathbf{x}_k) + \mathbf{P}$, where $\mathbf{P} \in \mathbb{R}^{N \times d_{\text{model}}}$ are learnable positional encodings that preserve the UPA spatial topology, and $\text{Linear}(\cdot): \mathbb{R}^{N \times 5} \rightarrow \mathbb{R}^{N \times d_{\text{model}}}$ is implemented as $\mathbf{x}_k \mathbf{W}_{\text{proj}} + \mathbf{b}_{\text{proj}}$ with weight matrix $\mathbf{W}_{\text{proj}} \in \mathbb{R}^{5 \times d_{\text{model}}}$ and bias vector $\mathbf{b}_{\text{proj}} \in \mathbb{R}^{d_{\text{model}}}$.

\subsubsection{Multi-Head Self-Attention}
The multi-head self-attention (MHSA) mechanism~\cite{vaswani2017attention} enables the model to capture non-local spatial correlations across the ELAA by allowing each antenna element to attend to all other elements simultaneously. Following~\cite{vaswani2017attention}, for layer $l$ with input $\mathbf{H}^{(l)} \in \mathbb{R}^{N \times d_{\text{model}}}$ and $h$ heads, query/key/value projections $\{\mathbf{W}_i^Q, \mathbf{W}_i^K, \mathbf{W}_i^V\}$ produce attention outputs:
\begin{equation}
  \text{Head}_i =
    \text{softmax}\!\left(
      \frac{\mathbf{Q}_i \mathbf{K}_i^T}{\sqrt{d_{\text{head}}}}
    \right)\mathbf{V}_i,
\end{equation}
which are concatenated and projected as:
\begin{equation}
  \resizebox{0.9\columnwidth}{!}{$\displaystyle
    \text{MHSA}(\mathbf{H}^{(l)}) =
      \text{Concat}(\text{Head}_1, \dots, \text{Head}_h)
      \mathbf{W}^O \in \mathbb{R}^{N \times d_{\text{model}}}.
  $}
\end{equation}
Here $d_{\text{head}}$ denotes the per-head key/query dimension, set to 64 in our experiments (see Table~\ref{tab:sim_params}). This mechanism dynamically weights each antenna element based on both the digital twin prior channel amplitude and the spherical wavefront curvature, prioritizing elements with constructive phase coherence.

\subsubsection{Output Projection and Training}
The learnable shared output parameters $\mathbf{W}_{\text{out}} \in \mathbb{C}^{1 \times d_{\text{model}}}$ and $\mathbf{b}_{\text{out}} \in \mathbb{C}$ are optimized by minimizing the mean squared error (MSE) between the estimated and ground-truth effective channels:
\begin{equation}
  \mathcal{L}_{\text{MSE}} =
    \frac{1}{K}\sum_{k=1}^{K}
    \| \hat{\mathbf{h}}_{\text{eff},k} - \mathbf{h}_{\text{eff},k} \|_2^2,
  \label{eq:loss_csi}
\end{equation}
where the ground-truth $\mathbf{h}_{\text{eff},k} \in \mathbb{C}^{N \times 1}$ is the full channel vector from the DeepVerse dataset including up to 5th-order reflected and scattered paths. Given the trained parameters, the channel estimate for user $k$ is produced by applying the shared complex-valued projection independently to each antenna token:
\begin{equation}
  \hat{\mathbf{h}}_{\text{eff},k} =
    \bigl[\mathbf{W}_{\text{out}}\mathbf{h}_{k,n}^{(L)}+\mathbf{b}_{\text{out}}\bigr]_{n=1}^{N}
    \in \mathbb{C}^{N \times 1},
  \label{eq:csi_output}
\end{equation}
where $L$ is the number of encoder layers and $\mathbf{h}_{k,n}^{(L)} \in \mathbb{R}^{d_{\text{model}}}$ is the $L$-th layer output for antenna element $n$. The shared projection preserves the UPA spatial structure and enables antenna-specific phase corrections, which is critical in the near-field regime where each antenna element observes a unique distance and phase. The complex-valued shared projection is applied independently to each antenna token, mapping the real-valued transformer output of antenna $n$ to the complex channel coefficient $[\hat{\mathbf{h}}_{\text{eff},k}]_n$. Under approximately Gaussian channel statistics, the MSE objective in~\eqref{eq:loss_csi} is an efficient approximation to maximum likelihood estimation~\cite{kay1993fundamentals}, enabling the transformer to learn optimal channel representations from limited training data.

\subsection{VIT-BASED BLOCKAGE PREDICTION}

The ViT blockage predictor is designed to identify motion patterns in sequential camera frames that precede physical blockage events, providing advance warning before link quality degrades. Unlike CNN classifiers that analyze each frame independently and thereby miss the temporal dynamics of approaching objects, or long short-term memory (LSTM) networks that model temporal sequences as flat vectors and thus discard intra-frame spatial structure, the ViT processes frames as sequences of spatial patches. This allows the self-attention mechanism to capture both within-frame spatial dependencies and across-frame temporal correlations simultaneously.

This joint spatiotemporal modeling capability is well-suited to the vehicular blockage prediction problem. In the Carla-Town5 environment, blockages are caused by large vehicles crossing between the BS and the UE, producing characteristic motion trajectories that extend across multiple spatial regions of the camera frame over several consecutive time steps. The patch-based representation ensures that the ViT attends to these structured motion cues, while stacking $F$ frames channel-wise within the effective channel dimension $C' = F \times C$ allows the network to correlate motion detected in early frames with the anticipated link states in later frames~\cite{ghassemi2025generative}. This mechanism directly enables the 769~ms advance warning window reported in our experiments, corresponding to the detection of approaching blockages five frames before they occur. Furthermore, the $16 \times 16$ patch embedding averages pixel-level noise over a spatial region before the attention layers operate, contributing to the robustness under camera blur observed in Section~\ref{sec:ablation}. The ViT processes each 10-frame sequence by aggregating the last ten camera frames (spanning approximately 1.5~s of history) to predict the link state in the upcoming 154~ms macro-step, ensuring temporal context extends well beyond a single macro-interval.

\subsubsection{Patch Embedding and Linear Projection}
The input consists of $F$ frames of resolution $H \times W$ with $C$ channels. Frames are stacked channel-wise to obtain effective channel dimension $C' = F \times C$, and each frame is divided into $N_p = HW/P^2$ non-overlapping patches of size $P \times P$. Prior to patching, each $960 \times 540$ pixel frame is resized to $720 \times 720$ pixels so that $720/16 = 45$ patches per side yield $N_p = 45^2 = 2025$ non-overlapping $16 \times 16$ patches. The patch sequence, prepended with a learnable classification token $\mathbf{x}_{\text{class}}$, is linearly embedded and added to positional encodings:
\begin{equation}
  \mathbf{z}_0 = [\mathbf{x}_{\text{class}};\;
                  \mathbf{x}^1_p \mathbf{E};\;
                  \mathbf{x}^2_p \mathbf{E};\;
                  \dots;\;
                  \mathbf{x}^{N_p}_p \mathbf{E}]
                + \mathbf{E}_{pos},
\end{equation}
where $\mathbf{E} \in \mathbb{R}^{(P^2 C') \times D}$ is the learnable patch embedding matrix and $\mathbf{E}_{pos} \in \mathbb{R}^{(N_p + 1) \times D}$ provides positional encodings that retain spatial information. The classification token $\mathbf{x}_{\text{class}}$ aggregates global scene context across all patches through the multi-head self-attention layers. Its final representation encodes whether the scene contains motion patterns indicative of an impending blockage, and is used exclusively for blockage probability prediction at the output head.

\subsubsection{Blockage Probability Output}
After $L_{\text{ViT}}$ transformer layers of MHSA and MLP blocks, the class token state $\mathbf{z}^{L_{\text{ViT}}}_{0}$ is passed through a sigmoid-activated MLP head~\cite{dosovitskiy2021image}:
\begin{equation}
  \hat{\mathbf{p}} =
    \delta\!\left(\text{MLP}\!\left(\text{LayerNorm}\!\left(
    \mathbf{z}^{L_{\text{ViT}}}_{0}\right)\right)\right),
\end{equation}
where $L_{\text{ViT}}$ denotes the number of ViT encoder blocks (distinct from $L$, the number of CSI transformer encoder layers defined in Section~\ref{sec:framework}-A) and $\hat{\mathbf{p}} = [\hat{p}_1, \dots, \hat{p}_K]^T \in [0,1]^K$ is the vector of per-user LoS link availability probabilities. Each element $\hat{p}_k$ represents the likelihood that the LoS link for user $k$ remains available within the next macro-interval $t'$, with $\hat{p}_k \approx 0$ indicating an impending blockage. These probabilities serve as the primary input to the DT-HDRL meta-controller, which uses them to decide in advance whether to activate the RIS-assisted path and which RIS units to switch on, rather than reacting after the link has already degraded. The ViT is trained by minimizing the binary cross-entropy (BCE) loss~\cite{dosovitskiy2021image}:
\begin{equation}
  \mathcal{L}_{\text{BCE}} = -\frac{1}{K}\sum_{k=1}^{K}
    \Big[
      b_k \log(\hat{p}_k + \epsilon)
      + (1-b_k)\log(1-\hat{p}_k + \epsilon)
    \Big],
  \label{eq:loss_blockage}
\end{equation}
where $b_k \in \{0,1\}$ is the ground-truth LoS availability label for user $k$ ($b_k = 1$ denotes LoS available, $b_k = 0$ denotes blocked), and $\epsilon = 10^{-7}$ ensures numerical stability.

\begin{algorithm*}[!htbp]
\caption{Dual-Transformer HDRL Training Procedure}
\label{alg:training}
\footnotesize
\SetAlgoLined
\SetInd{0.3em}{0.6em}
\setlength{\algomargin}{0.5em}
\textbf{Input:}
  Ray-tracing dataset $\mathcal{D}_{ray}$ (digital twin prior
  $\mathbf{h}_k^{\text{DT}}$ and ground-truth channels
  $\mathbf{h}_{\text{eff},k}$), visual dataset $\mathcal{D}_{vis}$,
  episodes $E$, macro-steps $T_{\text{macro}}$, micro-steps per macro
  $N_{\text{macro}}$, batch size $B$, learning rate $\eta$, discount
  factors $\gamma_l$, $\gamma_h$, target update rate
  $\tau_{\text{upd}}$, pre-training epochs $E_{\text{pre}}$\;
\textbf{Output:}
  Trained CSI transformer $\theta_{csi}$, ViT $\theta_{vit}$, HDRL
  policies $\pi_h$, $\pi_l$\;
\textbf{Initialize:}
  All networks with random weights; sub-controller replay buffer $\mathcal{B}_l
  \leftarrow \emptyset$; meta-controller trajectory buffer $\mathcal{T}_h
  \leftarrow \emptyset$; Ornstein-Uhlenbeck (OU)
  exploration noise with std $\sigma_o$ for sub-controller~\cite{lillicrap2016continuous}\;

\tcp{Phase 1: Supervised Pre-training (CSI Transformer and ViT)}
\For{pre-training epoch $e_{\text{pre}} = 1$ to $E_{\text{pre}}$}{
Sample batch
  $\{(\mathbf{h}^{\text{DT},(i)}_k,
  \mathbf{h}_{\text{eff}}^{(i)})\}_{i=1}^{B}$
  from $\mathcal{D}_{ray}$; construct fused input $\mathbf{x}_k^{(i)}$
  via~\eqref{eq:csi_input}\;
Forward pass:
  $\hat{\mathbf{h}}_{\text{eff}}^{(i)} \leftarrow
  \text{Transformer}_{\theta_{csi}}(\mathbf{x}_k^{(i)})$;
  compute $\mathcal{L}_{\text{MSE}}$ via~\eqref{eq:loss_csi};
  update $\theta_{csi} \leftarrow \theta_{csi} -
  \eta \nabla_{\theta_{csi}} \mathcal{L}_{\text{MSE}}$\;
Sample batch
  $\{(img^{(i)}, \mathbf{b}_{label}^{(i)})\}_{i=1}^{B}$
  from $\mathcal{D}_{vis}$\;
Forward pass:
  $\hat{\mathbf{p}}^{(i)} \leftarrow
  \text{ViT}_{\theta_{vit}}(img^{(i)})$;
  compute $\mathcal{L}_{\text{BCE}}$ via~\eqref{eq:loss_blockage};
  update $\theta_{vit} \leftarrow \theta_{vit} -
  \eta \nabla_{\theta_{vit}} \mathcal{L}_{\text{BCE}}$\;
}

\tcp{Phase 2: Hierarchical RL Training}
\For{episode $e=1$ to $E$}{
  Reset environment; observe initial state $s_0$\;
  \For{macro-step $t' = 1$ to $T_{\text{macro}}$}{
    Capture visual input; predict blockage probabilities:
      $\hat{\mathbf{p}}_{t'} \leftarrow
      \text{ViT}_{\theta_{vit}}(img_{t'})$\;
    Form meta-state:
      $s_{t'}^h = \{\hat{\mathbf{p}}_{t'}, \mathbf{L}_{t'}\}$
      where $\mathbf{L}_{t'} = \{\mathbf{u}_{k,t'}\}_{k=1}^K$\;
    Select subgoal
      $g_{t'} = \{\{c_{k,t'}\}_{k=1}^K,
      \{a_{r,t'}\}_{r \in \mathcal{R}}\} \sim
      \pi_h(s_{t'}^h)$
      where $c_{k,t'} \in \{0,1\}$ is the path selection for user $k$,
      and $a_{r,t'} \in \{0,1\}$ is the active/sleep decision for
      RIS $r$; enforce $\sum_{r \in \mathcal{R}} a_{r,t'} \ge \max_{k} c_{k,t'}$;
      initialize $R_{\text{meta}} \leftarrow 0$\;
    \For{micro-step $\tau = 1$ to $N_{\text{macro}}$}{
      \For{$k=1$ to $K$}{
        Query digital twin prior:
          $\mathbf{h}_k^{\text{DT}} \leftarrow
          \text{DigitalTwin}(\mathbf{u}_{k,\tau})$;
          compute geometric features
          $\{d_{k,n}^{\text{BD}},
          \vartheta_{k,n}^{\text{BD}},
          \psi_{k,n}^{\text{BD}}\}_{n=1}^N$
          via~\eqref{eq:csi_input};
          form fused input $\mathbf{x}_k \in \mathbb{R}^{N \times 5}$\;
        Estimate channel:
          $\hat{\mathbf{h}}_{\text{eff},k} \leftarrow
          \text{Transformer}_{\theta_{csi}}(\mathbf{x}_k)$\;
      }
      Form $\hat{\mathbf{H}}_{\text{eff},\tau} \leftarrow
        [\hat{\mathbf{h}}_{\text{eff},1}, \ldots,
        \hat{\mathbf{h}}_{\text{eff},K}] \in \mathbb{C}^{N \times K}$;
        form sub-state
        $s_\tau^l = \{\hat{\mathbf{H}}_{\text{eff},\tau}, g_{t'}\}$\;
      Select action
        $a_\tau^l = \{\mathbf{W}_\tau,
        \{\bm{\phi}_{r,\tau}\}_{r:\,a_{r,t'}=1}\} \leftarrow
        \pi_l(s_\tau^l) + \varepsilon_\tau^l$ (OU noise)\;
      Apply $\mathbf{W}_\tau$ at BS s.t.\
        $\|\mathbf{W}_\tau\|_F^2 \le P_{\text{max}}$;
        apply phase shifts $\{\bm{\phi}_{r,\tau}\}$ only at active
        RIS units where $a_{r,t'}=1$\;
      Compute reward $r_\tau^l = \sum_{k=1}^K \text{SE}_{k,\tau}$;
        observe $s_{\tau+1}^l$;
        store $(s_\tau^l, a_\tau^l, r_\tau^l, s_{\tau+1}^l)
        \rightarrow \mathcal{B}_l$\;
      \If{$|\mathcal{B}_l| \ge B$ and $\tau \bmod 4 = 0$}{
        Update sub-controller actor and critic via deep deterministic
        policy gradient (DDPG); soft-update target networks\;
      }
      $R_{\text{meta}} \leftarrow R_{\text{meta}} +
        \gamma_l^{\tau-1} r_\tau^l$\;
    }
    Append $(s_{t'}^h, g_{t'}, R_{\text{meta}}, s_{t'+1}^h)$
      to $\mathcal{T}_h$\;
    \If{$|\mathcal{T}_h| \ge B$}{
      Update meta-controller via PPO (clipped surrogate loss with
      independent Bernoulli action heads) using $\gamma_h$;
      $\mathcal{T}_h \leftarrow \emptyset$\;
    }
  }
}
\textbf{Return:} $\theta_{csi},\; \theta_{vit},\; \pi_h,\; \pi_l$
\end{algorithm*}

\subsection{TWO-TIMESCALE HDRL FOR JOINT OPTIMIZATION}

The HDRL agent consists of two interacting controllers that operate at distinct timescales and communicate through a shared subgoal mechanism. The Meta-controller operates at the slow timescale, indexed by macro-step $t'$, and issues a subgoal $g_{t'}$ at each macro-step that encodes both the transmission path for each user and the active/sleep state of each RIS unit. This subgoal remains fixed throughout the subsequent $N_{\text{macro}}$ micro-steps, during which the sub-controller operates at the fast timescale, indexed by micro-step $\tau$, and performs real-time optimization of beamfocusing and RIS phase shifts. The relationship between timescales is $T_{\text{macro}} = N_{\text{macro}} \cdot T_{\text{micro}}$, where $T_{\text{micro}} = 1$~ms is the sub-controller step duration and $N_{\text{macro}} = 154$ (see Table~\ref{tab:sim_params}). This separation is physically motivated: path selection and RIS scheduling depend on blockage predictions that evolve over seconds, whereas beamfocusing and phase shifts must track fast channel fading at the millisecond level. While alternating optimization (AO) and weighted minimum mean squared error (WMMSE) methods provide useful upper bounds, they require perfect channel state information and incur exponential complexity in the number of RIS phase-shift elements $M$. In contrast, the proposed DRL agent reduces the per-step inference complexity to 2.41 GFLOPs and operates directly on estimated CSI, making it suitable for millisecond-level scheduling in dynamic near-field environments.

Stable training of the dual-transformer architecture relies on several practical design choices.
(i)~Layer normalization in the ViT stabilizes patch embeddings.
(ii)~A shared complex-valued output projection in the CSI transformer preserves antenna-specific phase corrections.
(iii)~Gradient clipping at $1.0$ stabilizes the DDPG sub-controller.
(iv)~Warm-up epochs for transformer pre-training precede hierarchical RL.

The subgoal $g_{t'}$ from the meta-controller encodes two decisions for each macro-step. The path component $c_{k,t'} \in \{0,1\}$ specifies whether user $k$ is served via the direct LoS path ($c_{k,t'}=0$) or the RIS-assisted NLoS path ($c_{k,t'}=1$). The scheduling component $a_{r,t'} \in \{0,1\}$ specifies whether RIS $r$ is active or sleeping. The sub-controller receives $g_{t'}$ as part of its state and restricts its optimization to only the beamfocusing vectors and phase shifts of the active RIS units; it is unable to alter the path or scheduling decisions set by $g_{t'}$, even when instantaneous CSI might suggest a different routing would be beneficial. During back-propagation, gradients for the phase-shift outputs of sleeping RIS units ($a_{r,t'}=0$) are zero-masked so that the sub-controller network does not receive misleading updates for inactive hardware.

This strict one-way flow of authority from the meta-controller to the sub-controller avoids conflicts by design.
To ensure coherent operation across timescales, the meta-controller broadcasts the subgoal $g_{t'\!}$ to the sub-controller at the start of each macro-step, and the sub-controller reports the accumulated reward $R_{\text{meta}}$ back to the meta-controller at the end of the $N_{\text{macro}}$ micro-steps.
This closed-loop signal enables the meta-controller to learn which path and scheduling decisions yield the highest long-term spectral efficiency without interfering with the real-time beamfocusing adaptations of the sub-controller.

\textbf{Meta-controller (High-Level).} The meta-controller acts as the strategic planner over macro-steps $t' = 1, 2, \ldots, T_{\text{macro}}$. At each macro-step $t'$, it observes the ViT link availability probability vector $\hat{\mathbf{p}}_{t'}$ and the set of UE positions $\mathbf{L}_{t'}$, and issues the subgoal $g_{t'}$ that simultaneously determines, for each user $k$, which transmission path to use and, for each RIS $r$, whether the unit should be active or asleep. When any user is assigned to the RIS-assisted path ($c_{k,t'}=1$ for some $k$), at least one RIS unit is required to be active, as enforced by the conditional constraint~(C4) of Problem~\eqref{eq:overall_opt}. This proactive path and scheduling decision is issued $N_{\text{macro}}$ micro-steps in advance of the actual blockage event, exploiting the 769~ms advance warning provided by the ViT.

\textbf{Sub-controller (Low-Level).} The sub-controller acts as the real-time optimizer over micro-steps $\tau = 1, 2, \ldots, N_{\text{macro}}$ within each macro-step $t'$. At each micro-step $\tau$, it receives the instantaneous estimated CSI matrix $\hat{\mathbf{H}}_{\text{eff},\tau} \in \mathbb{C}^{N \times K}$ together with the fixed subgoal $g_{t'}$ issued by the meta-controller for the current macro-step. Using these inputs, the sub-controller determines the optimal BS beamfocusing matrix $\mathbf{W}_\tau$ and the phase-shift vectors $\{\bm{\phi}_{r,\tau}\}$ for all active RIS units (those with $a_{r,t'}=1$), maximizing the sum SE within the routing and scheduling strategy set by $g_{t'}$. The update frequency of every 4 micro-steps balances sample efficiency with training stability, a common practice in DDPG implementations.

Each controller is modeled as a Markov decision process (MDP) $\langle \mathcal{S}, \mathcal{A}, \mathcal{R}, \gamma \rangle$, defined as follows.

\textbf{1) Meta-Controller MDP.}
\begin{itemize}
  \item \textbf{State.}
    $s_{t'}^h = \{\hat{\mathbf{p}}_{t'}, \mathbf{L}_{t'}\}$, where
    $\hat{\mathbf{p}}_{t'} \in [0,1]^K$ is the ViT link availability
    probability vector and $\mathbf{L}_{t'} =
    \{\mathbf{u}_{k,t'}\}_{k=1}^K$ is the set of UE positions at
    macro-step $t'$.
  \item \textbf{Action.}
    $g_{t'} = \{\{c_{k,t'}\}_{k=1}^K,
    \{a_{r,t'}\}_{r \in \mathcal{R}}\} \in \{0,1\}^{K +
    |\mathcal{R}|}$, where $c_{k,t'} \in \{0,1\}$ is the path
    selection for user $k$ and $a_{r,t'} \in \{0,1\}$ is the
    active/sleep decision for RIS $r$.
  \item \textbf{Reward.}
    $R_{t'}^h = \sum_{\tau=1}^{N_{\text{macro}}} \gamma_l^{\tau-1}
    r_\tau^l$, the sub-controller rewards accumulated within
    macro-step $t'$, discounted by $\gamma_l$.
    The meta-controller temporal-difference update across macro-steps
    uses $\gamma_h$.
\end{itemize}

\textbf{2) Sub-Controller MDP.}
\begin{itemize}
  \item \textbf{State.}
    $s_\tau^l = \{\hat{\mathbf{H}}_{\text{eff},\tau}, g_{t'}\}$,
    where $\hat{\mathbf{H}}_{\text{eff},\tau} \in \mathbb{C}^{N
    \times K}$ is the estimated effective channel matrix at micro-step
    $\tau$ and $g_{t'}$ is the subgoal fixed for the current
    macro-step.
  \item \textbf{Action.}
    $a_\tau^l = \{\mathbf{W}_\tau,
    \{\bm{\phi}_{r,\tau}\}_{r:\,a_{r,t'}=1}\}$, jointly encoding the
    BS beamfocusing matrix and the phase-shift vectors for active RIS
    units only, where $\bm{\phi}_{r,\tau} = [\phi_{1,r,\tau}, \ldots, \phi_{M,r,\tau}]^T \in [0, 2\pi)^M$ and $\bm{\Theta}_{r,\tau} = \text{diag}(e^{j\bm{\phi}_{r,\tau}})$.
  \item \textbf{Reward.}
    $r_\tau^l = \sum_{k=1}^K \text{SE}_{k,\tau} = \sum_{k=1}^K
    \log_2(1 + \text{SINR}_{k,\tau})$.
\end{itemize}

The sub-controller is trained with DDPG~\cite{lillicrap2016continuous} for continuous beamfocusing and phase-shift optimization, while the meta-controller is trained with PPO~\cite{schulman2017proximal} for discrete path-selection and RIS active/sleep decisions. Since the meta-controller outputs strictly binary decisions, we employ proximal policy optimization (PPO) rather than continuous DDPG. The meta-controller policy comprises independent Bernoulli action heads for each path-selection variable $c_{k,t'}$ and each RIS scheduling variable $a_{r,t'}$. At each macro-step, the policy probabilities are sampled to produce the subgoal $g_{t'}$. PPO updates use the clipped surrogate objective computed over a trajectory buffer of recent macro-step transitions, which provides stable on-policy learning for discrete combinatorial actions. Separate experience replay and trajectory buffers ensure independent and stable learning at each hierarchy level. The meta-controller learns to associate visual blockage threats with long-term strategic routing and scheduling decisions, while the sub-controller focuses on real-time beamfocusing adaptation within the fixed strategic context. This hierarchical decomposition builds upon recent advances in multi-timescale RL for wireless networks~\cite{zhou2022hierarchical,habib2025llm}.

\subsection{COMPLEXITY ANALYSIS}

Standard least-squares (LS) or minimum mean square error (MMSE) estimators for RIS-assisted systems scale as $\mathcal{O}(KMN)$ or $\mathcal{O}(KM^2)$~\cite{hu2021two}, becoming a bottleneck for large intelligent surfaces and growing proportionally with the number of RIS units. The proposed CSI transformer achieves $\mathcal{O}(L(N^2 + Nd_{\text{model}})K)$ per inference, independent of $M$. The HDRL agent inference cost is $\mathcal{O}(N_{\text{layer}} d_{\text{hidden}}^2 + NKd_{\text{hidden}} + |\mathcal{R}|d_{\text{hidden}})$ for the meta-controller, which selects path and scheduling decisions, and $\mathcal{O}(N_{\text{layer}} d_{\text{hidden}}^2 + NKd_{\text{hidden}} + |\mathcal{R}|Md_{\text{hidden}})$ for the sub-controller, where the last term accounts for the phase-shift action heads of active RIS units. The ViT runs once per macro-interval and its cost is amortized over $N_{\text{macro}}$ micro-steps, so the per-transmission time interval (TTI) overhead is determined solely by the CSI transformer and HDRL agent. Full GFLOPs figures and measured latencies for the complete system configuration ($N=1024$, $M=100$, $K=10$) are reported in Section~\ref{sec:complexity}.

\section{NUMERICAL RESULTS AND DISCUSSION}
\label{sec:results}

\subsection{DEEPVERSE 6G CARLA-TOWN5 DATASET}

We employ the Carla-Town5 scenario from the DeepVerse 6G dataset~\cite{demirhan2025deepverse}, a publicly available benchmark designed for machine-learning-driven 6G research. Channel realizations are generated using the Remcom Wireless InSite ray-tracing engine with up to 5th-order reflections and diffuse scattering. All channel realizations are evaluated at a carrier frequency of $f_c = 3.5$~GHz to study near-field beamfocusing in a sub-6 GHz macro-cellular scenario, which is relevant for early 6G outdoor deployments.

Multiple RGB cameras are co-located with the BS to cover the surrounding road segments, enabling joint training of DT-HDRL. Ground-truth blockage labels for ViT training are extracted from the ray-tracing simulator's LoS visibility data. The binary LoS availability label $b_k \in \{0, 1\}$ for UE $k$ at each time step is derived as:
\begin{equation}
  b_k =
  \begin{cases}
1, & \text{if } PL_{\text{LoS},k} < PL_{\text{th}}, \\ 0, & \text{otherwise,}
  \end{cases}
\end{equation}
where $PL_{\text{LoS},k}$ is the measured LoS path loss and $PL_{\text{th}}$ is the threshold specified in Table~\ref{tab:sim_params}. A label of $b_k = 1$ indicates that the LoS path loss is below the threshold, confirming an available LoS link, while $b_k = 0$ indicates a blocked link. The ViT is therefore trained to predict LoS availability, with a low output value $\hat{p}_k \approx 0$ signaling an impending blockage. We partition the dataset into 3,200 training and 800 test samples (80/20 split), with a held-out 10\% of training used as a validation set during hyperparameter tuning, ensuring strict separation between splits.

\subsection{SIMULATION SETUP}

The system comprises $K = 10$ single-antenna vehicular UEs distributed within a $150 \times 100 \times 1.5$~m$^3$ outdoor road region. The BS employs a $32 \times 32$ UPA ($N = 1024$ antennas) at $[0, 0, 0]^T$~m. Two RIS units are deployed: RIS~1 with $M=100$ elements at $[40, 0, 10]^T$~m and RIS~2 with $M=100$ elements at $[-40, 0, 10]^T$~m, each independently subject to active/sleep scheduling by the meta-controller at each macro-step $t'$. Key parameters are summarized in Table~\ref{tab:sim_params}.

\begin{table}[h!]
\centering
\caption{Simulation Parameters}
\label{tab:sim_params}
\small
\setlength{\tabcolsep}{2pt}
\scriptsize
\begin{tabular}{|l|l|}
\hline
\textbf{Parameter} & \textbf{Value} \\ \hline
\multicolumn{2}{|c|}{\textbf{System Configuration}} \\ \hline
BS Antennas ($N = N_y \times N_z$) & $1024 = 32 \times 32$ \\
Number of RIS units ($|\mathcal{R}|$) & 2 \\
RIS Elements per unit ($M = M_y \times M_z$) & $100 = 10 \times 10$ \\
Carrier Frequency ($f_c$) & 3.5 GHz \\
Wavelength ($\lambda$) & 85.7 mm \\
System Bandwidth & 100 MHz \\
Noise Power Spectral Density ($\sigma^2$) & $-94$ dBm \\
SNR Range & $-10$ to 50 dB \\
Number of UEs ($K$) & 10 \\
UE Antennas & 1 \\
BS / RIS Element Spacing & $\lambda/2$ \\
BS Position ($\mathbf{u}_{\text{BS}}$) & $[0, 0, 0]^T$ m \\
RIS~1 Position ($\mathbf{u}_{\text{RIS},1}$) & $[40, 0, 10]^T$ m \\
RIS~2 Position ($\mathbf{u}_{\text{RIS},2}$) & $[-40, 0, 10]^T$ m \\
RIS Active/Sleep Scheduling & Per unit, per macro-step $t'$ \\
UE Region (outdoor road segment) & $150 \times 100 \times 1.5$ m$^3$ \\
Rayleigh Distance ($Z_R$, 2D UPA diagonal) & $\approx$82 m \\
Max Transmit Power ($P_{\text{max}}$) & 35 dBm \\
Min SE per User ($\text{SE}_{\text{min}}$) & 1 bps/Hz \\ \hline
\multicolumn{2}{|c|}{\textbf{Visual Data Collection}} \\ \hline
Camera Frame Rate & 6.5 samples/s \\
Image Resolution & $960 \times 540$ pixels \\
Resized Input for Patching & $720 \times 720$ pixels \\
Frames per Sequence ($F$) & 10 \\
Blockage Threshold ($PL_{\text{th}}$) & Noise floor + 10 dB \\ \hline
\multicolumn{2}{|c|}{\textbf{Timescale Configuration}} \\ \hline
Meta-Controller Step Duration ($T_{\text{macro}}$) & 154 ms \\
Sub-Controller Step Duration ($T_{\text{micro}}$) & 1 ms   \\
Micro-steps per Macro ($N_{\text{macro}}$)
  & 154 ($T_{\text{macro}} = 154\,T_{\text{micro}}$) \\ \hline
\multicolumn{2}{|c|}{\textbf{Training Configuration}} \\ \hline
Number of Training Episodes ($E$) & 5000 \\
Macro-steps per Episode ($T_{\text{macro}}$) & 100 \\
Training Samples ($N_{\text{data}}$) & 4,000 (80/20 split) \\
Training Epochs & 1000 (CSI/ViT) \\
Learning Rate ($\eta$) & $10^{-4}$ \\
Batch Size ($B$) & 64 (CSI), 256 (HDRL) \\
Optimizer & Adam \\
BCE Stability Constant ($\epsilon$) & $10^{-7}$ \\ \hline
\multicolumn{2}{|c|}{\textbf{HDRL Parameters}} \\ \hline
Sub-Controller Replay Buffer $|\mathcal{B}_l|$ & $10^5$ \\
Meta-Controller Trajectory Buffer $|\mathcal{T}_h|$ & $10^4$ \\
Sub Discount $\gamma_l$ / Meta Discount $\gamma_h$ & 0.99 / 0.9 \\
Sub-Controller Target Update Rate $\tau_{\text{upd}}$ & 0.001 \\
Sub-Controller Exploration Noise $\sigma_o$ & 0.3 (OU process) \\
Sub-Controller Update Frequency & Every 4 micro-steps \\
Meta-Controller Algorithm & PPO (clipped surrogate) \\
PPO Clip Parameter $\epsilon_{\text{PPO}}$ & 0.2 \\
PPO Epochs per Update & 4 \\ \hline
\multicolumn{2}{|c|}{\textbf{CSI Transformer Architecture}} \\ \hline
Input Feature Dimension
  & 5 (DT prior re/im + $d^{\text{BD}}$,
    $\vartheta^{\text{BD}}$, $\psi^{\text{BD}}$) \\
Embedding Dimension ($d_{\text{model}}$) & 512 \\
Encoder Layers ($L$) / Heads ($h$) & 6 / 8 \\
Per-Head Dimension ($d_{\text{head}}$) & 64 \\
Activation Function & ReLU \\ \hline
\multicolumn{2}{|c|}{\textbf{ViT Architecture}} \\ \hline
Embedding Dimension ($D$) & 768 \\
Encoder Blocks ($L_{\text{ViT}}$) / Heads & 12 / 12 \\
Patch Size ($P$) / Patches ($N_p$) & $16{\times}16$ / 2025 \\
MLP Hidden Layers & [512, 256, 128] \\
Activation Function & ReLU \\ \hline
\end{tabular}
\end{table}

\subsection{BASELINE METHODS}

\subsubsection{CSI Estimation Baselines}
\textbf{(1) CNN.} Three convolutional layers (64/128/256 filters); this baseline receives the same fused digital-twin-prior plus geometric-correction input as the proposed transformer but is limited by local receptive fields and is unable to capture non-local spherical-wave correlations~\cite{wang2022transformer}. \textbf{(2) MLP.} A fully connected network with the same 5-feature fused input, isolating the contribution of the self-attention structure. \textbf{(3) LS.} Conventional pilot-based estimator; scales as $\mathcal{O}(1/\text{SNR})$, where SNR denotes signal-to-noise ratio. \textbf{(4) MMSE Oracle.} Assumes perfect second-order channel statistics; provides a lower bound on MSE.

\subsubsection{Blockage Prediction Baselines}
\textbf{(5) Vanilla Transformer.} A transformer encoder without patch-based processing. \textbf{(6) CNN Classifier.} Frame-by-frame CNN; no temporal modeling. \textbf{(7) LSTM.} Temporal sequence model; no spatial patch structure.

\subsubsection{Resource Optimization Baselines}
\textbf{(8) DRL w/o RIS.} Single-level DDPG, BS beamfocusing only, no hierarchical decomposition or RIS. \textbf{(9) DRL w/ RIS.} Single-level DDPG, joint RIS scheduling, beamfocusing, and phase-shift optimization, no timescale decomposition. \textbf{(10) HDRL w/o RIS.} Hierarchical DDPG with dual-timescale decomposition but without RIS assistance. \textbf{(11) AO.} Alternates between zero-forcing beamfocusing and RIS phase optimization with \emph{perfect} CSI; an SE upper bound. \textbf{(12) WMMSE.} Weighted minimum mean squared error (WMMSE) iterative convex approximation with perfect CSI; a second SE upper bound. \textbf{(13) FC-DNN.} Fully connected deep neural network (FC-DNN) agent with the same state and action dimensions as the sub-controller, isolating the benefit of hierarchical decomposition.

Note that the proposed DT-HDRL employs a hybrid PPO/DDPG hierarchy (PPO for the discrete meta-controller, DDPG for the continuous sub-controller), whereas all RL baselines use uniform DDPG.

\subsection{CONVERGENCE AND CSI ESTIMATION ACCURACY}

We first evaluate the CSI transformer in isolation to separate the contributions of the digital twin prior fusion and the self-attention architecture.

\begin{figure}[t]
  \centering
  \includegraphics[width=\columnwidth]{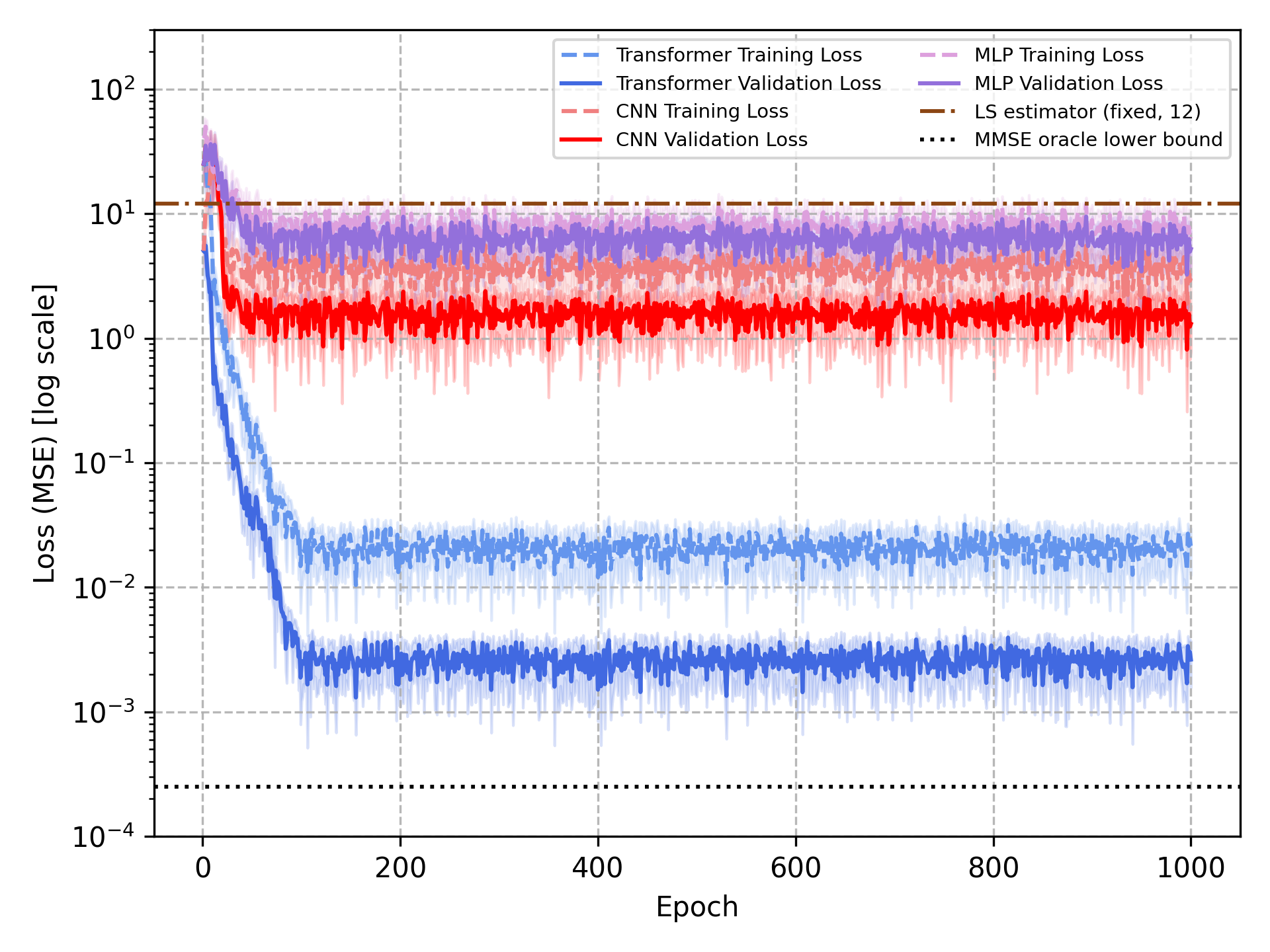}
  \caption{Training and validation MSE loss over 1000 epochs (mean $\pm$ 95\% CI, 10 runs).
    The proposed transformer (digital-twin-prior + geometric correction) achieves 73\% lower MSE than CNN and 80\% lower than MLP.
    LS and MMSE oracle are shown as horizontal references.}
  \label{fig:csi_training_convergence}
\end{figure}

\textbf{Training Convergence.} Fig.~\ref{fig:csi_training_convergence} compares training and validation MSE loss over 1000 epochs for all four estimators (LS and MMSE oracle appear as horizontal references). The proposed transformer achieves the lowest final MSE, reducing it by 73\% relative to CNN and 80\% relative to MLP. The full transformer's validation loss approaches within one order of magnitude of the MMSE oracle, confirming near-optimal use of the available digital twin prior and geometric features.

\begin{figure}[t]
  \centering
  \includegraphics[width=\columnwidth]{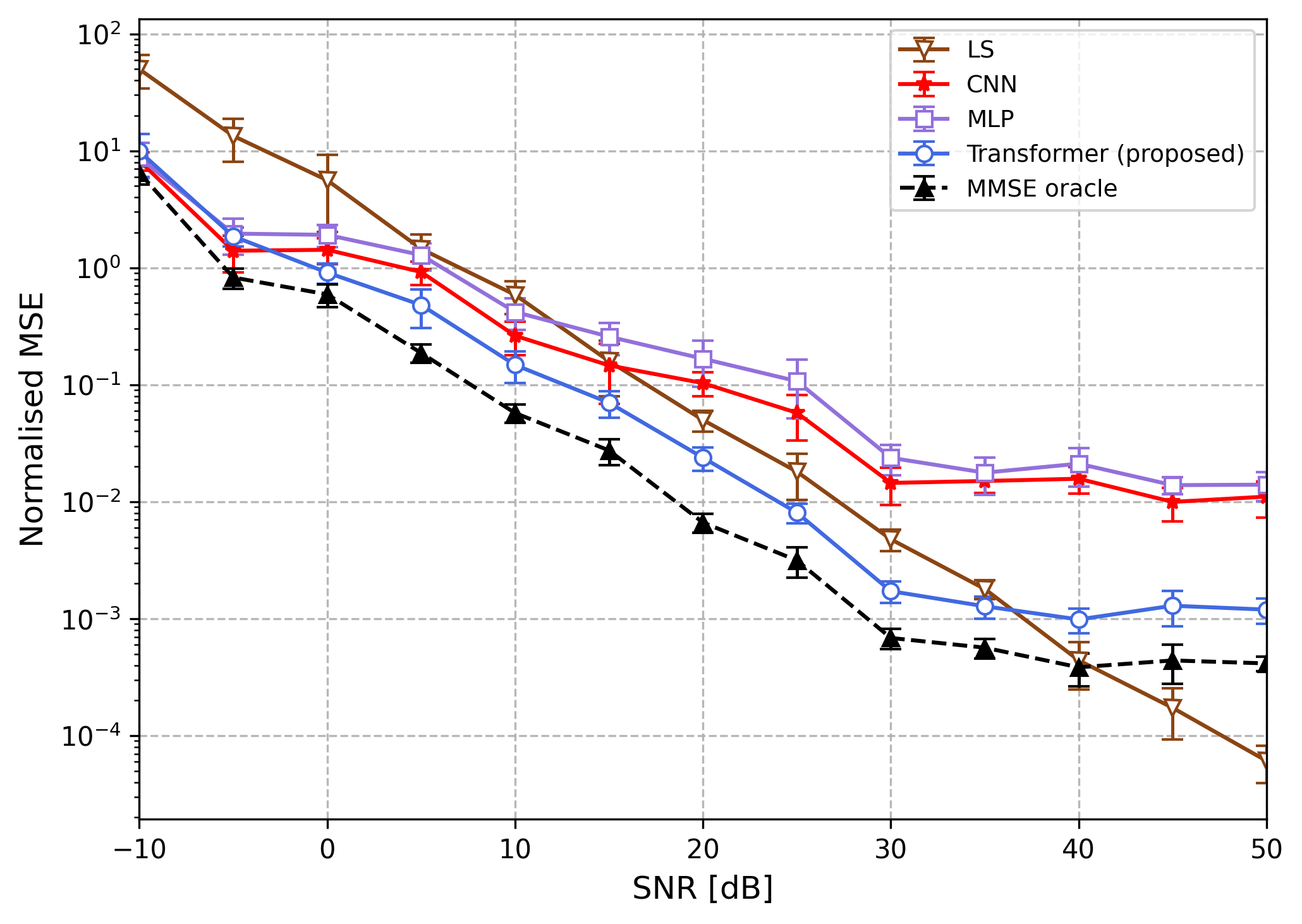}
  \caption{Normalized MSE versus SNR (mean $\pm$ std, 8 independent realizations).
    The proposed transformer maintains the lowest error and approaches the MMSE oracle at high SNR.}
  \label{fig:csi_snr}
\end{figure}

\textbf{SNR Performance.} Fig.~\ref{fig:csi_snr} plots normalized MSE versus SNR for all methods, with error bars representing the standard deviation over eight independent channel realizations. The LS estimator degrades as a slope-$(-1)$ line on the log scale. CNN and MLP plateau at moderate SNR due to limited spatial modelling capacity, while the proposed transformer continues to improve and approaches the MMSE oracle at high SNR. Error bars confirm statistical significance across independent realizations.

\subsection{BLOCKAGE PREDICTION PERFORMANCE}

\begin{figure}[t]
  \centering
  \includegraphics[width=\columnwidth]{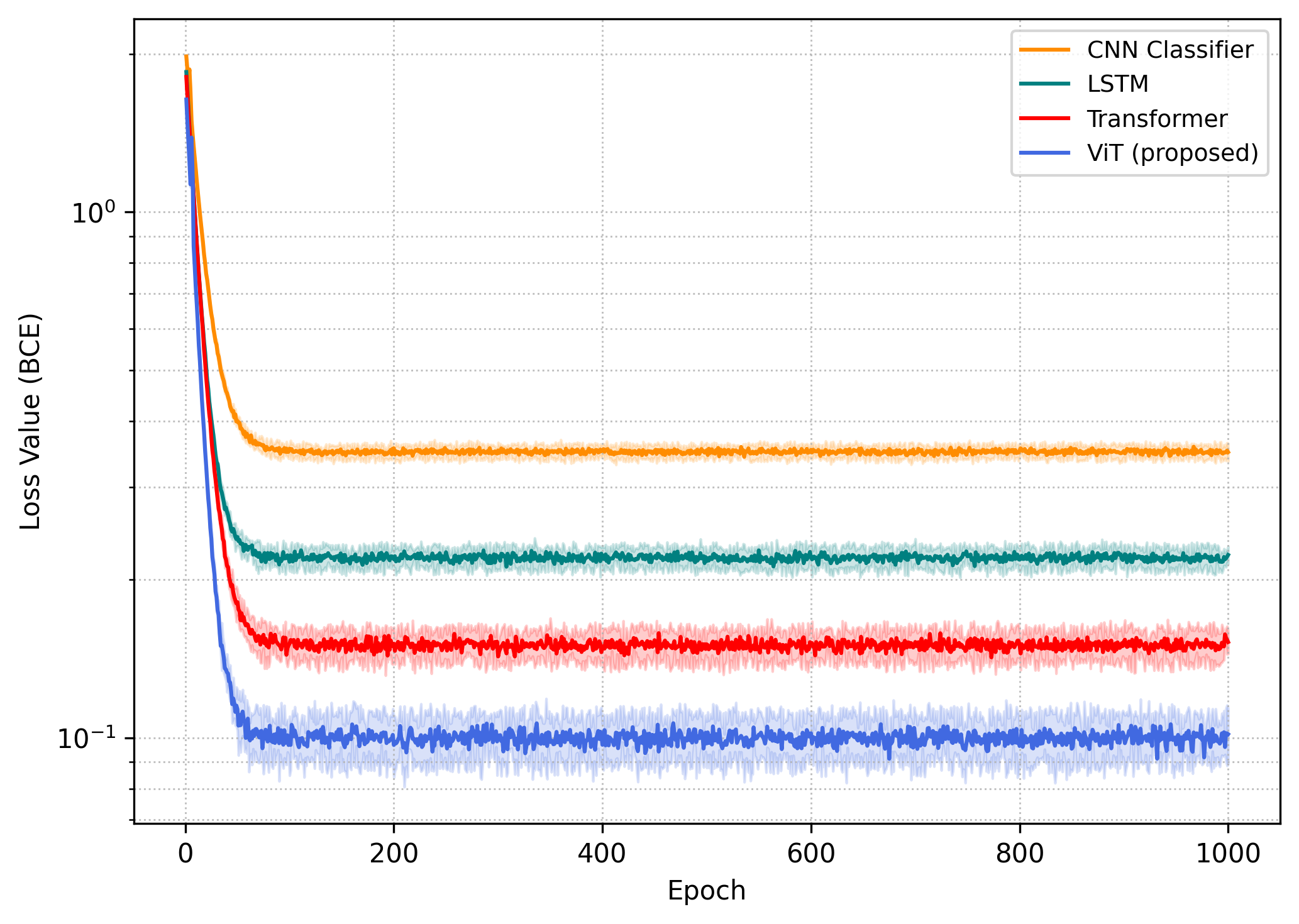}
  \caption{Training BCE loss over 1000 epochs (mean $\pm$ 95\% CI).
    The ViT achieves the lowest final loss; the CNN Classifier plateaus earliest due to the absence of temporal modeling.}
  \label{fig:vit_training}
\end{figure}

Fig.~\ref{fig:vit_training} compares BCE loss convergence over 1000 epochs for all four methods. The CNN Classifier converges fastest but plateaus highest because it processes each frame independently, missing temporal context. The LSTM captures temporal dynamics and outperforms the CNN Classifier, but lacks the spatial patch structure suited to image-level motion cues. The vanilla Transformer improves further through global attention, while the ViT combination of patch embedding and positional encoding achieves the lowest and most stable final loss. As reported in Table~\ref{tab:blockage_metrics}, the ViT achieves an F1-score of 0.92, correctly predicting blockages 5 frames before occurrence ($\approx 769$~ms). An F1-score of 0.92 implies that 8\% of strategic mode switches are mistimed; these errors manifest as transient SE dips of roughly 3\% during the 154~ms macro-interval, after which the sub-controller compensates via fast-timescale beam adaptation.

\begin{table}[t]
\centering
\caption{Blockage Prediction Performance}
\label{tab:blockage_metrics}
\begin{tabular}{|l|c|c|c|}
\hline
\textbf{Method} & \textbf{Precision} & \textbf{Recall}
  & \textbf{F1-Score} \\ \hline
Proposed ViT         & 0.94 & 0.90 & \textbf{0.92} \\
Transformer Baseline & 0.86 & 0.82 & 0.84 \\
LSTM                 & 0.81 & 0.78 & 0.80 \\
CNN Classifier       & 0.76 & 0.73 & 0.74 \\ \hline
\end{tabular}
\end{table}

\subsection{SE PERFORMANCE}

\textbf{Dynamic Evolution.}

\begin{figure}[t]
  \centering
  \includegraphics[width=\columnwidth]{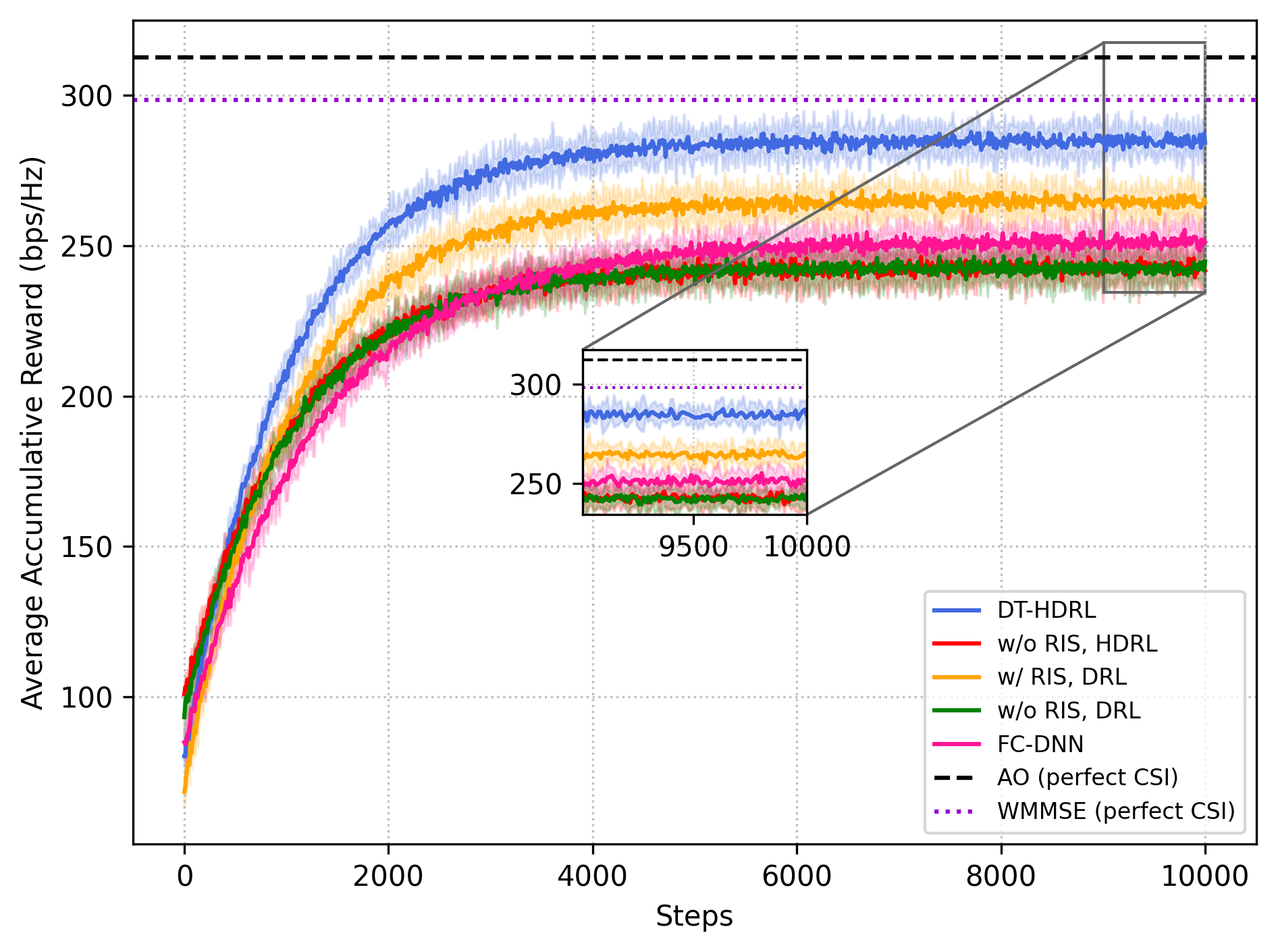}
  \caption{Moving-average sum SE during training for all RL baselines.
    FC-DNN converges more slowly and to a lower value than DT-HDRL.
    AO and WMMSE are shown as constant upper-bound references.}
  \label{fig:se_steps}
\end{figure}

\begin{figure}[t]
  \centering
  \includegraphics[width=\columnwidth]{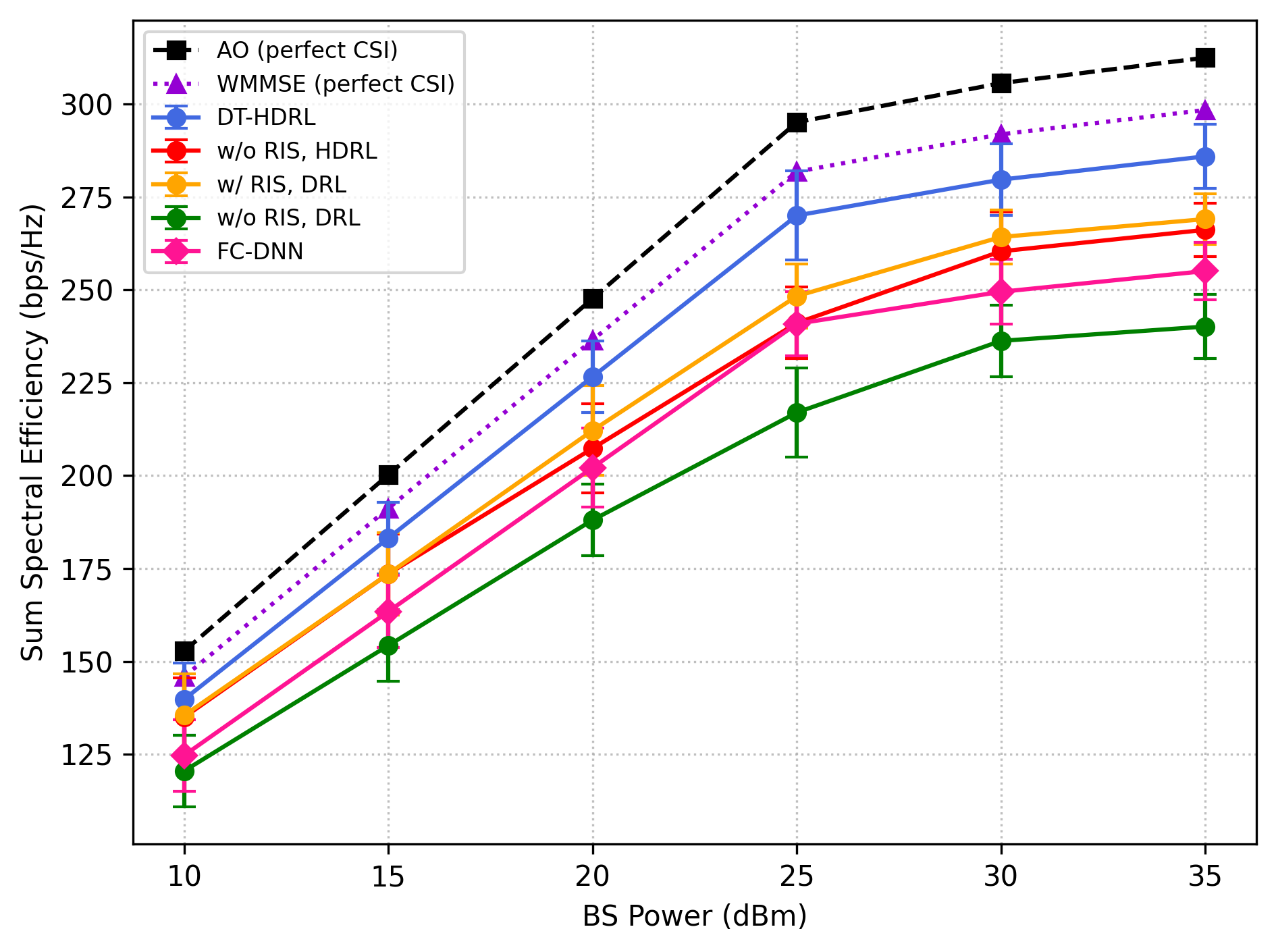}
  \caption{Converged sum SE versus BS transmit power ($N=1024$, $M=100$ per RIS, $|\mathcal{R}|=2$).
    AO and WMMSE with perfect CSI define the upper envelope.
    DT-HDRL reaches 91.5\% of the AO bound at $P_{\max}=35$~dBm.}
  \label{fig:se_power}
\end{figure}

Fig.~\ref{fig:se_steps} tracks the moving-average sum SE during the exploratory training phase. DT-HDRL converges to approximately 285.9~bps/Hz, an 18.0\% gain over the HDRL baseline without RIS, which saturates at approximately 242.7~bps/Hz. This gain arises from two complementary sources: the meta-controller stabilizes learning by abstracting blockage dynamics, and the active/sleep scheduling allows the sub-controller to selectively engage only the RIS units that provide constructive NLoS coverage, reducing inter-RIS interference and improving the effective SNR at blocked UEs. Steady-state values are reported in Figs.~\ref{fig:se_power}--\ref{fig:se_ris}.

\textbf{Impact of Transmit Power.} Fig.~\ref{fig:se_power} shows SE versus transmit power. At $P_{\text{max}} = 35$~dBm, DT-HDRL outperforms the non-RIS DRL baseline by $\approx 9.5$~bps/Hz and reaches 91.5\% of the AO upper bound, confirming that the HDRL agent effectively leverages blockage predictions to activate the appropriate RIS units and configure their phase shifts before link degradation.

\subsection{ABLATION AND ROBUSTNESS ANALYSIS}
\label{sec:ablation}

Table~\ref{tab:ablation} reports the ablation results for the default configuration ($N=1024$, $M=100$, $|\mathcal{R}|=2$, $K=10$). Removing any single component degrades sum SE by 7--22\%, with the distance-aware feature $d_{k,n}^{\text{BD}}$ contributing most (21.8\%), confirming the importance of each design element for near-field beamfocusing. Removing the ViT blockage predictor reduces SE by 14.3\% and increases outage probability by a factor of 3.2, because the reactive scheme switches paths only after the link has already failed. Removing the hierarchical decomposition reduces SE by 7.4\%, showing that the two-timescale structure provides a meaningful advantage beyond RIS alone. DT-HDRL reaches 91.5\% of the AO bound and 95.8\% of the WMMSE bound despite operating with estimated CSI, predicted blockages, and learned RIS active/sleep decisions. To further validate robustness, we separately evaluate the effects of imperfect CSI and UE location errors in the following subsections. The results confirm that DT-HDRL maintains superior performance when the CSI transformer operates on noisy inputs and when the geometric features are derived from perturbed locations, since the hierarchical controller compensates via fast-timescale SINR feedback.

\begin{table}[t]
\centering
\caption{Ablation Study ($N=1024$, $M=100$, $|\mathcal{R}|=2$,
  $K=10$)}
\label{tab:ablation}
\begin{tabular}{|l|c|}
\hline
\textbf{Configuration} & \textbf{Sum SE (bps/Hz)} \\ \hline
Full DT-HDRL (proposed)
  & 285.9 \\
Without distance feature $d_{k,n}^{\text{BD}}$ in CSI transformer
  & 223.4 \\
Without ViT blockage prediction (reactive)
  & 245.1 \\
Without hierarchical control (single-level DRL)
  & 264.7 \\
Without RIS (HDRL w/o RIS)
  & 242.7 \\
FC-DNN agent (non-hierarchical)
  & 251.3 \\
AO (perfect CSI, upper bound)
  & 312.5 \\
WMMSE (perfect CSI, upper bound)
  & 298.4 \\ \hline
\end{tabular}
\end{table}

\subsubsection{Robustness to Imperfect CSI and Location Errors}
\begin{figure}[t]
  \centering
  \includegraphics[width=\columnwidth]{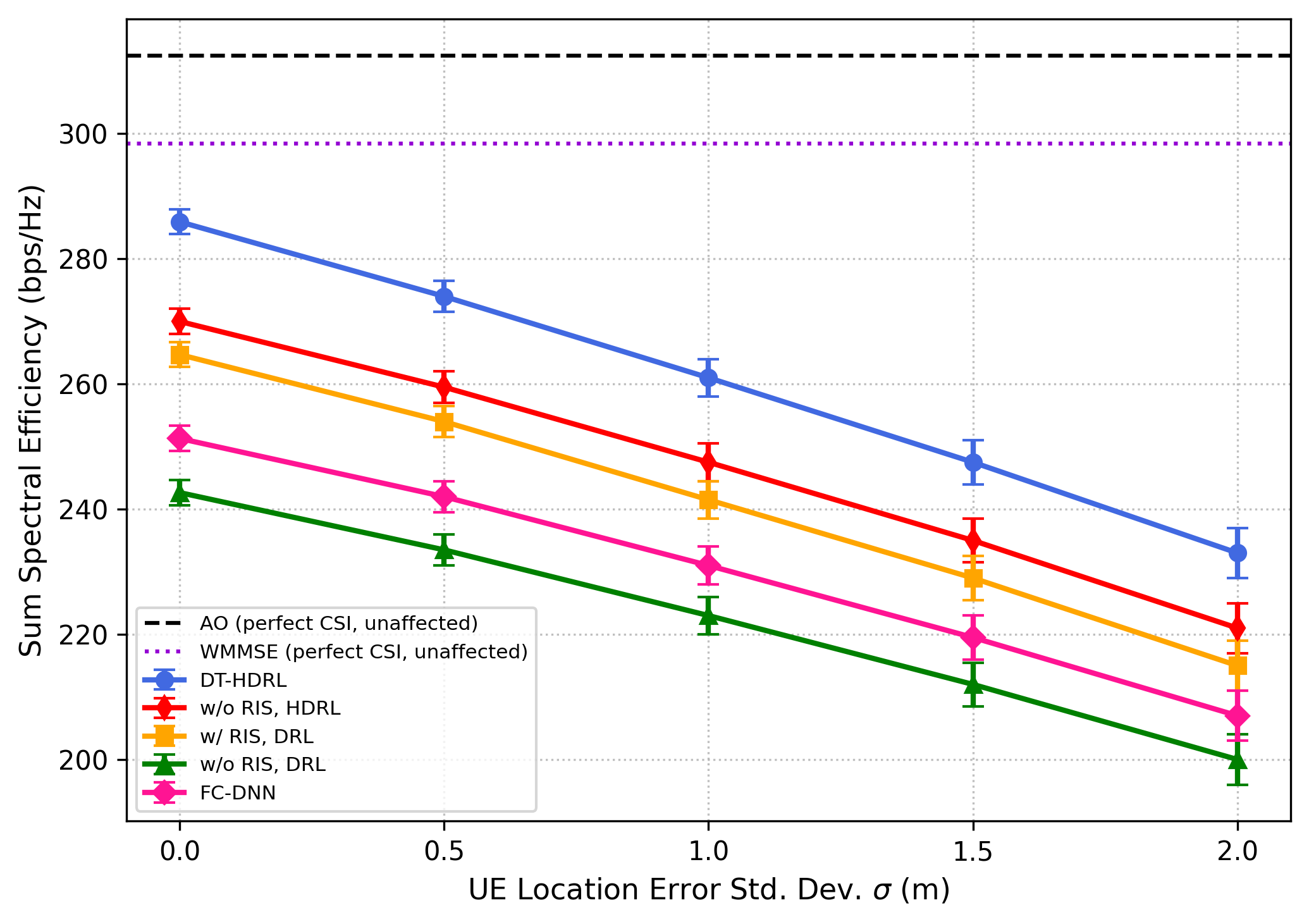}
  \caption{Sum SE versus UE location error standard deviation $\sigma_{\text{loc}}$.
    DT-HDRL degrades most gracefully among all learning-based methods.}
  \label{fig:robust_loc}
\end{figure}

Fig.~\ref{fig:robust_loc} evaluates robustness by introducing Gaussian noise to the UE location estimates used for CSI feature extraction. At a location error standard deviation of $\sigma_{\text{loc}} = 0.5$~m, sum SE degrades by only 4.2\% (to 274.0~bps/Hz). Even with $\sigma_{\text{loc}} = 1$~m, SE remains above 260~bps/Hz, owing to the HDRL ability to adapt beamfocusing and RIS active/sleep decisions based on received SINR feedback. The digital-twin-prior is also resilient to small location errors because the dominant ray paths shift only slightly when the query position changes by sub-meter amounts within the near-field regime. AO and WMMSE use perfect CSI and therefore appear as constant horizontal references; this comparison bounds the performance gap between estimated and perfect CSI operation.

\begin{figure}[t]
  \centering
  \includegraphics[width=\columnwidth]{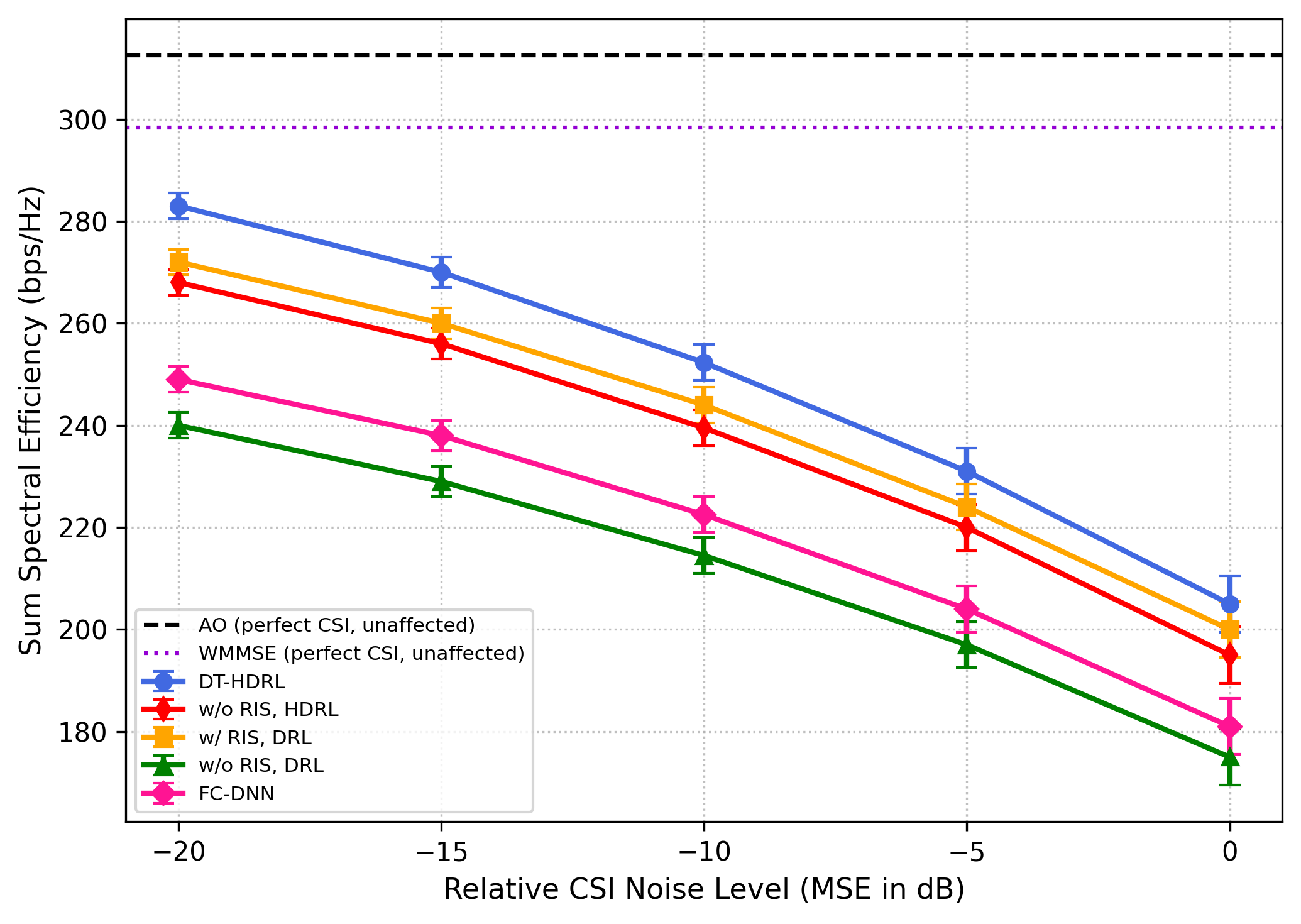}
  \caption{Sum SE versus additive CSI noise level (MSE in dB relative to the true channel).
    DT-HDRL outperforms all learning-based baselines at every noise level.}
  \label{fig:robust_csi}
\end{figure}

When the estimated CSI is corrupted by additive noise (MSE $= -10$~dB relative to the true channel), Fig.~\ref{fig:robust_csi} shows that SE drops to 252.3~bps/Hz, still outperforming all learning-based baselines.

\subsubsection{Robustness of ViT to Sensing Noise}
\begin{figure}[t]
  \centering
  \includegraphics[width=\columnwidth]{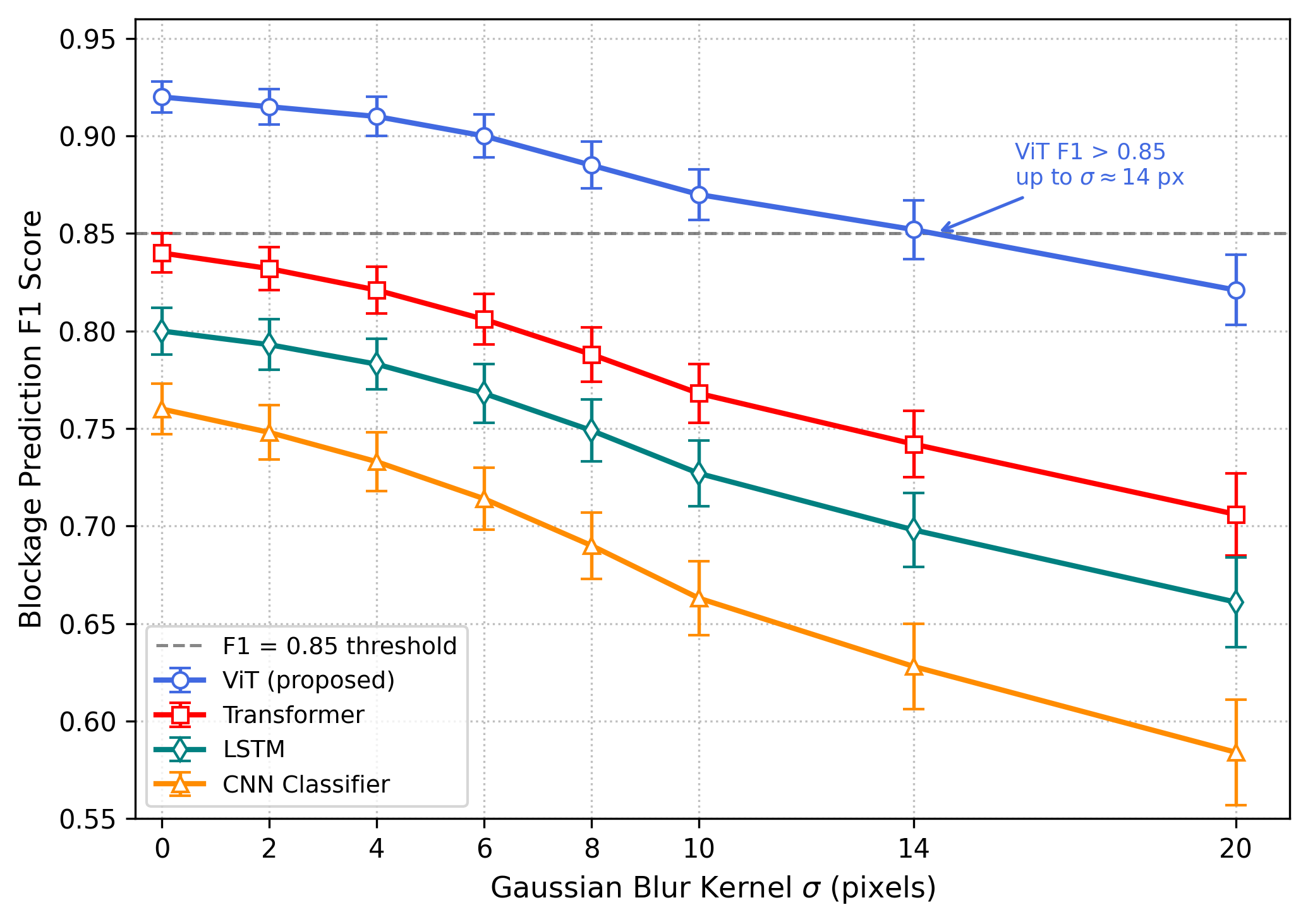}
  \caption{Blockage prediction F1-score versus Gaussian blur severity (kernel standard deviation $\sigma_{\text{blur}}$ in pixels of a $960\times540$ image).
    The ViT maintains F1~$> 0.85$ until $\sigma_{\text{blur}} \approx 14$~pixels, outperforming Transformer, LSTM, and CNN Classifier at all distortion levels.}
  \label{fig:vit_blur}
\end{figure}

Fig.~\ref{fig:vit_blur} evaluates ViT robustness under Gaussian blur of increasing severity (kernel standard deviation $\sigma_{\text{blur}}$ from 0 to 20 pixels of the $960\times540$ image). The ViT maintains F1~$> 0.85$ until $\sigma_{\text{blur}} \approx 14$~pixels, outperforming all baselines at every blur level. This resilience stems from the $16 \times 16$ patch embedding, which averages over a spatial region and thereby attenuates pixel-level noise before the attention layers operate.

\subsection{ARRAY SCALING ANALYSIS}

\begin{figure}[t]
  \centering
  \includegraphics[width=\columnwidth]{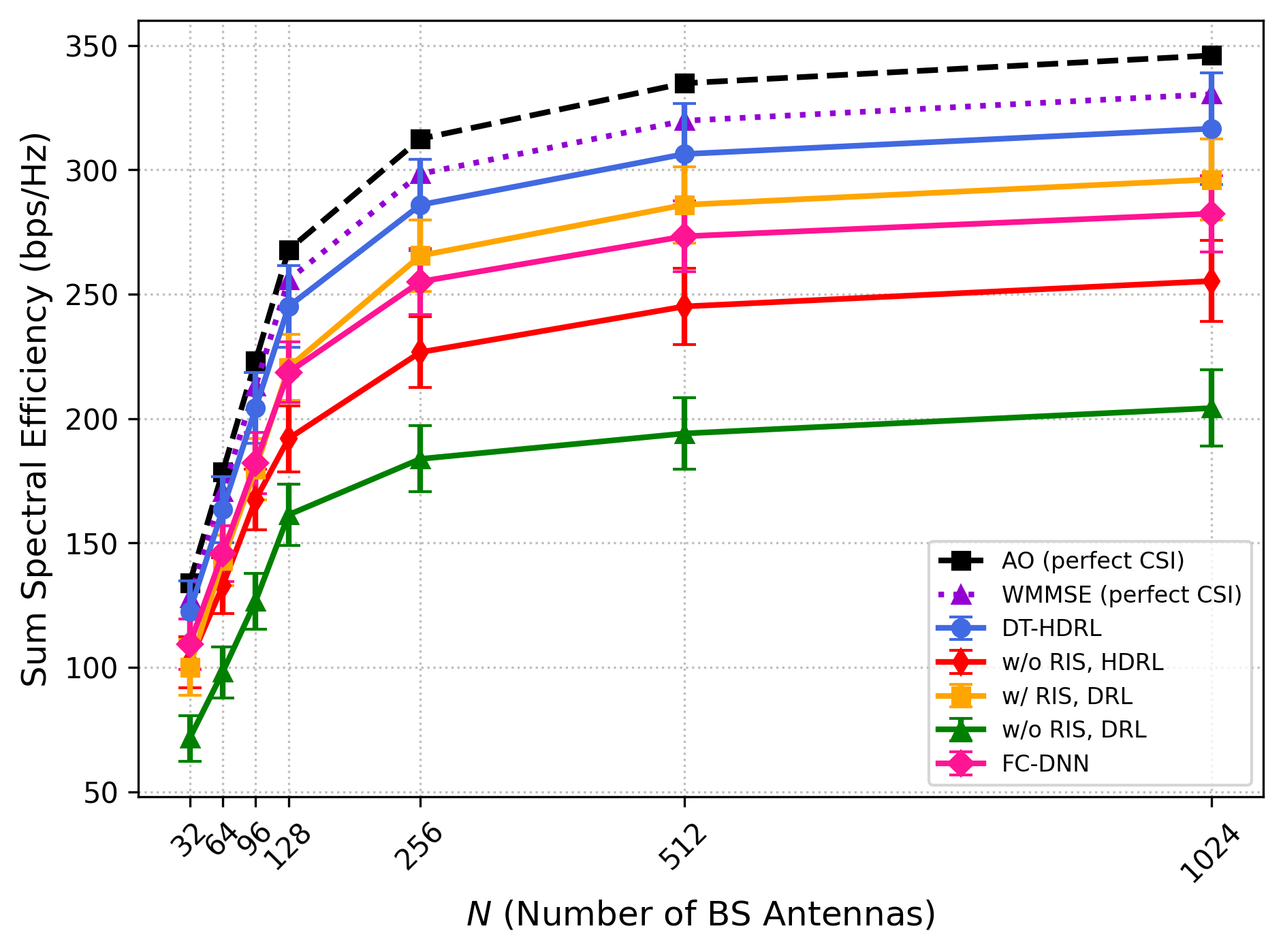}
  \caption{Average SE versus number of BS antennas $N$ (mean $\pm$ std).
    The performance gap over FC-DNN and single-level DRL widens with larger arrays.}
  \label{fig:se_n}
\end{figure}

\textbf{BS Antenna Scaling.} Fig.~\ref{fig:se_n} shows SE versus the number of BS antennas $N$ from 32 to 1024. The performance gap between DT-HDRL and all baselines widens with larger arrays, because the digital-twin-prior-fused transformer-based CSI estimator scales more effectively to high-dimensional spaces than standard estimators. The carrier frequency and element spacing are held constant across all $N$ values, so varying $N$ directly scales the aperture and Rayleigh distance. For small $N$ (e.g., $N=64$ configured as $8\times8$), the Rayleigh distance shrinks to approximately 4.2~m and the system may operate in the far-field; a rigorous study of this regime is left for future work.

\begin{figure}[t]
  \centering
  \includegraphics[width=\columnwidth]{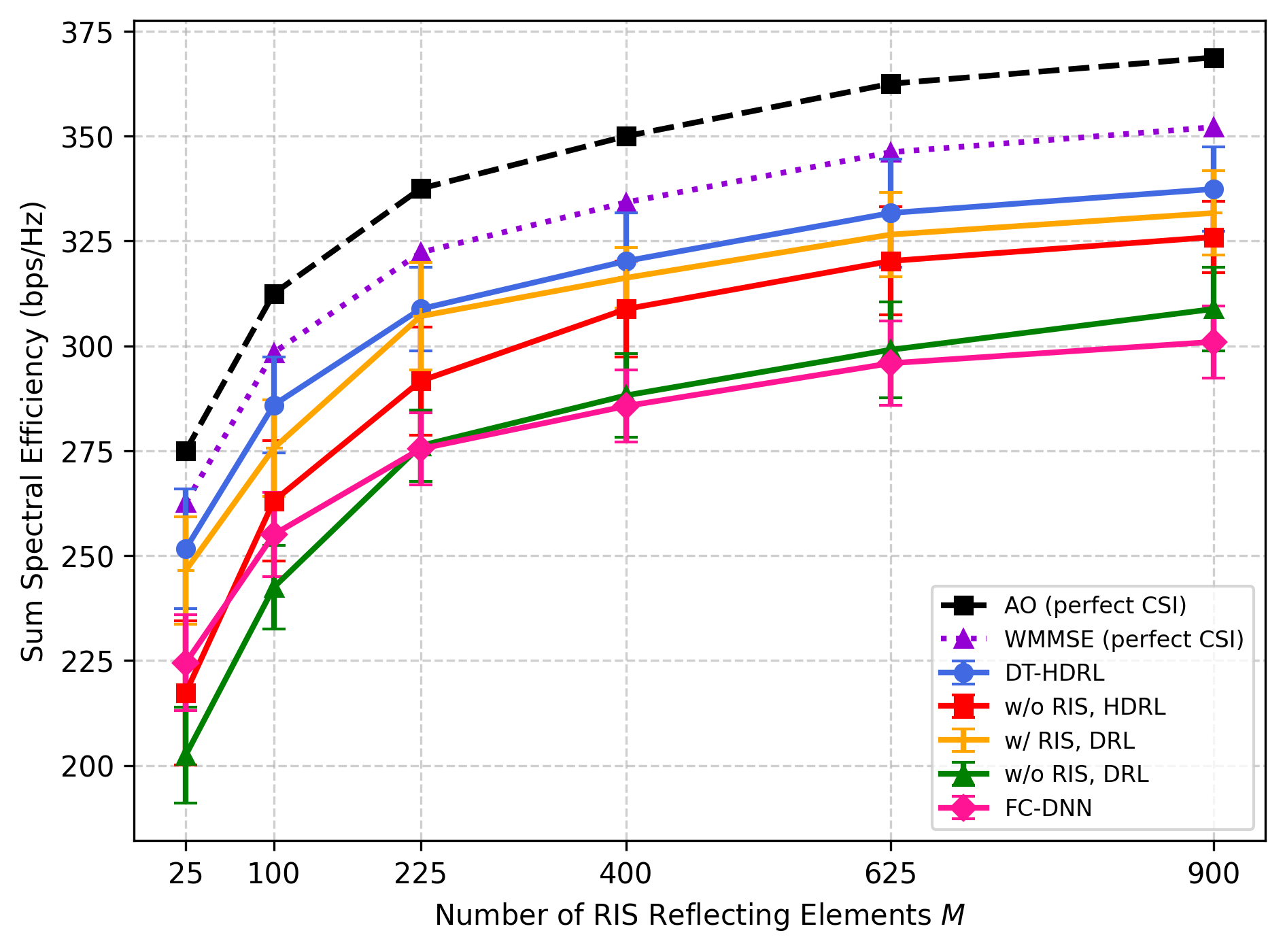}
  \caption{Average SE versus RIS reflecting elements $M$ per unit ($|\mathcal{R}|=2$).
    Performance saturates beyond $M=625$.
    DT-HDRL consistently exceeds 91\% of the AO upper bound.}
  \label{fig:se_ris}
\end{figure}

\textbf{RIS Element Scaling.} Fig.~\ref{fig:se_ris} illustrates the impact of RIS size $M$ (applied identically to both RIS units). Increasing $M$ from 25 to 625 yields significant gains, but performance saturates beyond $M=625$, indicating a physical limit imposed by the spatial correlation of the near-field channel. Adding elements beyond the coherence aperture yields diminishing returns, providing a design guideline for RIS dimensioning. Specifically, the RIS aperture exceeds the spatial coherence length of the near-field channel beyond this point, causing correlated phase shifts that do not increase the effective degrees of freedom.

\subsection{COMPUTATIONAL COMPLEXITY VALIDATION}
\label{sec:complexity}

All measurements correspond to the full system configuration ($N=1024$, $M=100$, $|\mathcal{R}|=2$, $K=10$) on an Intel Core i7-11370H CPU, 64~GB RAM, and NVIDIA A100-40~GB GPU. Compared with AO and WMMSE methods that require perfect CSI and suffer exponential complexity in $M$, the DRL inference provides sub-millisecond latency, handles estimated CSI, and scales gracefully to large arrays. The CSI transformer requires 0.85~GFLOPs per forward pass. The digital twin prior query to the ray-tracing engine is performed offline and cached; its latency does not impact real-time inference. The HDRL agent requires 1.56~GFLOPs per inference, including the active/sleep scheduling head. On the NVIDIA A100-40~GB GPU with FP16 precision, the combined CSI transformer and HDRL agent inference completes in less than 0.15~ms per TTI, well within the 1~ms coherence time. The ViT blockage predictor requires approximately 271.5~GFLOPs per forward pass, comprising 247.6~GFLOPs for the twelve transformer layers ($N_p = 2025$, $D = 768$) and 23.9~GFLOPs for the patch embedding ($C' = F \times C = 30$, $P = 16$). At 40\% hardware utilization on the A100-40~GB GPU with FP16 precision, the ViT forward pass completes in approximately 2.2~ms, running once per 154~ms macro-interval and therefore incurring negligible per-TTI overhead. Results are summarized in Table~\ref{tab:complexity_val}.

\begin{table}[t]
\centering
\caption{Computational Complexity per TTI ($N=1024$, $M=100$,
  $|\mathcal{R}|=2$, $K=10$)}
\label{tab:complexity_val}
\footnotesize
\setlength{\tabcolsep}{2pt}
\resizebox{\columnwidth}{!}{%
\begin{tabular}{|l|c|c|}
\hline
\textbf{Component} & \textbf{GFLOPs} & \textbf{Latency (ms)} \\ \hline
CSI Transformer ($N = 1024$, $K = 10$)
  & 0.85 & $<$0.05 \\
HDRL Agent ($N=1024$, $|\mathcal{R}|M=200$, $K=10$,
  incl.\ scheduling)
  & 1.56 & $<$0.10 \\
\textbf{Total per micro-step (excl.\ ViT)}
  & \textbf{2.41} & \textbf{$<$0.15} \\
ViT Blockage Predictor ($N_p = 2025$, $L_{\text{ViT}} = 12$, $C' = 30$)
  & 271.5 & $\approx$2.2 \\
\textbf{Total per macro-frame}
  & \textbf{273.9} & \textbf{$<$2.4} \\ \hline
\end{tabular}%
}
\end{table}

\section{CONCLUSION}
\label{sec:conclusion}

DT-HDRL demonstrates that information quality, rather than algorithmic optimality, is the binding constraint in RIS-assisted near-field systems envisioned for 6G. Accurate, timely CSI and proactive blockage awareness are both necessary, and neither alone is sufficient. By aligning two specialized transformer models with the two natural information timescales of the problem, specifically millisecond CSI updates and second-level blockage evolution, and by exploiting ray-tracing priors to reduce data requirements, the framework achieves 91.5\% of the perfect-CSI AO bound and an 18.0\% SE improvement over single-timescale baselines, while operating entirely on estimated, noisy inputs with learned RIS active/sleep scheduling, with the ViT link availability predictor providing a 769~ms advance warning at an F1-score of 0.92.

The proposed framework leverages a ray-tracing digital twin that closely approximates real-world propagation geometry, thereby bridging the gap between simulated training and practical deployment. Consequently, the DT-HDRL agent can be directly transferred to physical testbeds with only minor calibration of the geometric correction branch.

Three concrete limitations shape future directions. First, RIS element scaling saturates beyond $M = 625$ elements due to the spatial coherence aperture of the near-field channel, motivating coherence-aperture-aware RIS placement. Second, for small arrays ($N \leq 64$) the Rayleigh distance falls below 5~m and the near-field assumption breaks down, motivating a hybrid near-field/far-field controller that switches regimes based on estimated user distance. Third, the ViT 8\% misprediction rate causes transient SE dips during mistimed mode switches, motivating fault-tolerant or confidence-weighted subgoal mechanisms in the meta-controller. Further extensions include hybrid beamfocusing architectures, wideband beam squint mitigation, V2X communication scenarios, high-speed mobility-aware HDRL designs, and electromagnetic exposure constraints enforced via reward shaping.

\section*{ACKNOWLEDGMENT}

This work has been supported by the NSERC Canada Research Chairs program, MITACS, and Ericsson. We also wish to commemorate our co-author, Han, whose exceptional commitment and meaningful contributions strengthened this work. Her warmth and enthusiasm will be sincerely missed.

\bibliographystyle{IEEEtran}
\bibliography{bibliography}


\begin{IEEEbiography}[{\includegraphics[width=1in,height=1.25in,clip,keepaspectratio]{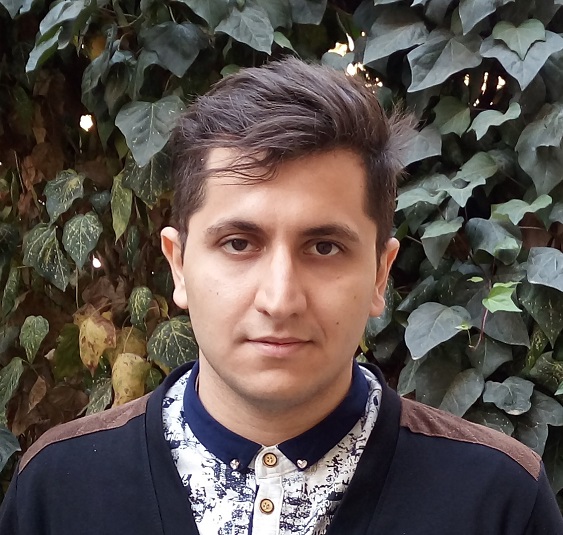}}]{Mohammad Ghassemi}
received the B.Sc. and M.Sc. degrees in electrical engineering and computer science. He is currently pursuing the Ph.D. degree at the School of Electrical Engineering and Computer Science, University of Ottawa. His research interests include machine learning for wireless communications, foundation models, and autonomous network management for next-generation mobile networks.
\end{IEEEbiography}

\begin{IEEEbiography}[{\includegraphics[width=1in,height=1.25in,clip,keepaspectratio]{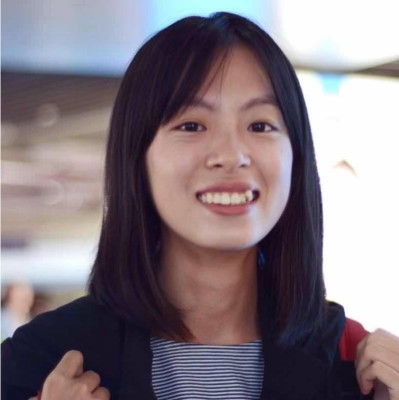}}]{Han Zhang}
received the B.Sc. and Ph.D. degrees from the School of Electrical Engineering and Computer Science, University of Ottawa. Her research focuses on the application of large language models to wireless network management, including network intrusion detection and AI-native communication systems.
\end{IEEEbiography}

\begin{IEEEbiography}[{\includegraphics[width=1in,height=1.25in,clip,keepaspectratio]{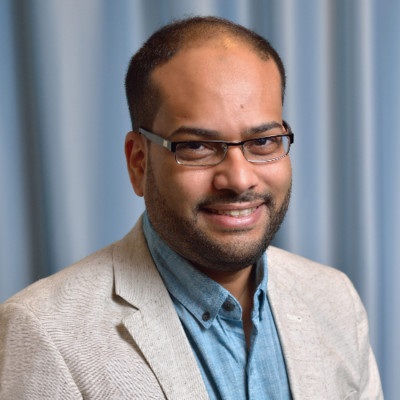}}]{Akram Bin Sediq}
(Senior Member, IEEE) received the Ph.D. degree in electrical engineering from Carleton University. He is currently a Principal AI/ML Developer at Ericsson and an Adjunct Research Professor at Carleton University. His research interests include radio resource management and machine learning for wireless communications.
\end{IEEEbiography}

\begin{IEEEbiography}[{\includegraphics[width=1in,height=1.25in,clip,keepaspectratio]{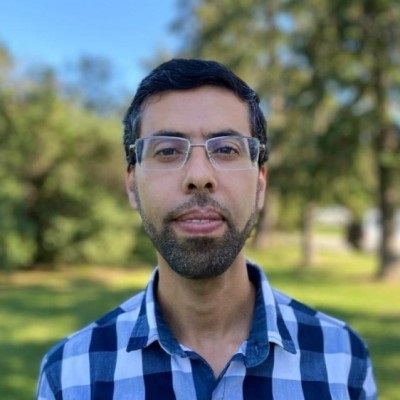}}]{Ali Afana}
(Senior Member, IEEE) received the Ph.D. degree in electrical engineering from Concordia University. He is currently a 5G Systems Developer at Ericsson. His research interests include wireless communications, signal processing, and machine learning for next-generation communication systems.
\end{IEEEbiography}

\begin{IEEEbiography}[{\includegraphics[width=1in,height=1.25in,clip,keepaspectratio]{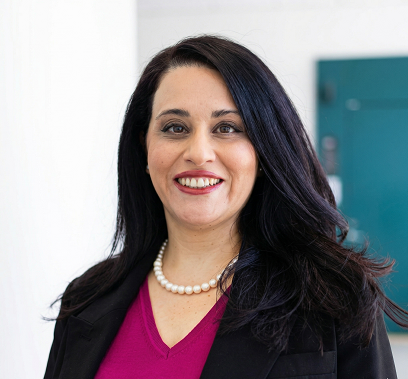}}]{Melike Erol-Kantarci}
(Fellow, IEEE) is currently a Canada Research Chair in AI-enabled Next-Generation Wireless Networks and a Full Professor with the School of Electrical Engineering and Computer Science, University of Ottawa. She is also the Strategic Product Manager for AI in RAN with Ericsson, and the Founding Director of the Networked Systems and Communications Research (NETCORE) Laboratory. She has received several awards, including the IEEE Women in Engineering 2025 Outstanding Achievement Award, and was an IEEE ComSoc Distinguished Lecturer from 2020 to 2023. She has served as the General Chair and Technical Program Chair for many international conferences, and is on the Editorial Board of IEEE Transactions on Communications, IEEE Transactions on Network Science and Engineering, and IEEE Networking Letters. She is the co-editor of four books on explainable AI in networks, smart grids, and intelligent transportation. She has over 250 peer-reviewed publications and has delivered more than 80 keynotes, plenary talks, and tutorials. Her research interests include AI-enabled wireless networks, 5G/6G communications, smart grids, and the Internet of Things.
\end{IEEEbiography}

\vfill\pagebreak

\end{document}